\documentclass[12pt]{article}
\usepackage{amsbsy}
\usepackage{amsfonts}
\usepackage{amsmath}
\usepackage{amssymb}
\usepackage{apacite}
\usepackage{setspace}
\usepackage{graphicx}
\usepackage{bm,multicol}
\usepackage[english]{babel}
\usepackage{natbib}
\usepackage[T1]{fontenc}
\usepackage[utf8]{inputenc}
\usepackage{authblk}
\usepackage{xcolor}

\newtheorem{lemma}{Lemma}

\newcommand{\newc}{\newcommand}
\newc{\N}{\mbox{N}}
\newc{\1}{\bf{1}}

\begin{document}
\title{Default Bayesian Model Selection of Constrained Multivariate Normal Linear Models\footnote{A revised version of this article has been accepted for publication in \textit{Multivariate Behavioral Research}, published by Taylor and Francis.}}

\author{J. Mulder, H. Hoijtink, \& X. Gu}


\maketitle

\begin{abstract}
The multivariate normal linear model is one of the most widely employed models for statistical inference in applied research. Special cases include (multivariate) $t$ testing, (M)AN(C)OVA, (multivariate) multiple regression, and repeated measures analysis. Statistical procedures for model selection where the models may have equality and order constraints on the model parameters of interest are limited however. This paper presents a default Bayes factor for this model selection problem. The default Bayes factor is based on generalized fractional Bayes methodology where different fractions are used for different observations and where the default prior is centered on the boundary of the constrained space under investigation. First, the method is fully automatic and therefore can be applied when prior information is weak or completely unavailable. Second, using group specific fractions, the same amount of information is used from each group resulting in a minimally informative default prior having a matrix Cauchy distribution, resulting in a consistent default Bayes factor. Third, numerical computation can be done using parallelization which makes it computationally cheap. Fourth, the evidence can be updated in a relatively simple manner when observing new data. Fifth, the selection criterion can be applied relatively straightforwardly in the presence of missing data that are missing at random. Applications for the social and behavioral sciences are used for illustration.
\end{abstract}


\section{Introduction}
The multivariate normal linear model is one of the most widely used statistical models for applied research. Special cases include multivariate $t$ testing, multivariate analysis of variance, multivariate multiple regression, and repeated measures analysis. In applied research these models are used to test scientific expectations which are typically formulated using competing equality and order constraints on the parameters of interest \citep{VanWell:2008,Schoot:2011c,Kluytmans:2012,Braeken:2015,Vrinten:2016,vanSchie:2016,DeJong:2017,Flore:2019,Digge:2019,Zondervan:2019}. This can be formalized as a model selection problem of $T$ constrained models of the form,
\begin{equation}
M_t:\textbf{R}_{t,E}\bm\theta=\textbf{r}_{t,E}~\&~\textbf{R}_{t,O}\bm\theta>\textbf{r}_{t,O},
\label{Mt}
\end{equation}
where $\bm\theta$ is a vector of (adjusted) means and regression coefficients of interest, and $[\textbf{R}_{t,E}|\textbf{r}_{t,E}]$ and $[\textbf{R}_{t,O}|\textbf{r}_{t,O}]$ are augmented matrices containing the coefficients of the $r_{t,E}$ equality constraints and $r_{t,O}$ order constraints under $M_t$, respectively, for $t=1,\ldots,T$. The goal is then to determine which model receives most evidence from the data at hand.

First it is important to note that classical significance tests are not particularly suitable for this problem as we would require to specify a central (null) model, which we would then reject or not given the observed data. Furthermore classical p values are not available for testing multiple nonnested models such as $M_1:\beta_1=\beta_2=\beta_3$ versus $M_2:\beta_1=\beta_2<\beta_3$ versus $M_3:\beta_1<\beta_2=\beta_3$ versus $M_4:\text{neither }H_1,~H_2,~\text{nor }H_3$. Furthermore information criteria such as the AIC, BIC, or DIC are also not suitable for this model selection problem as these methods quantify the complexity of a model based on the number of free parameters. This number however is ill-defined for models containing order constraints.

For this reason we focus on the Bayes factor, the Bayesian criterion for model selection and hypothesis testing \cite{Jeffreys}, given its proven effectiveness for selecting among competing order constrained models \citep{Hoijtink:2011}. A well-known property of the Bayes factor is that it can be very sensitive to the chosen prior distribution of the parameters that are tested \citep{Kass:1995,DeSantis:1997}. Arbitrary prior specification should therefore be avoided. Prior specification for the parameters under all the $T$ separate models based on prior beliefs can be a challenging and time-consuming endeavour however \citep{Berger:2006}. Furthermore noninformative improper priors can also not be used as the Bayes factor would depend on undefined constants. For this reason there has been an extensive development of so-called default or automatic Bayes factors where a small subset of the data is used for prior specification \citep[e.g.][and the references therein]{Spiegelhalter:1982,OHagan:1995,Berger:1996,Berger:1999,Moreno:1998,Berger:2004,Klugkist:2005,Mulder:2009,Rouder:2009,Klugkist:2010,Gu:2014,Gu:2017,Consonni:2017,BoingMessing:2017,Mulder:2018,MulderOlsson:2019}.

So far however there has been very little literature on default Bayes factors under the multivariate normal linear model despite the wide spread applicability of the model. An exception is the default Bayes factor proposed in \cite{Mulder:2010} which was implemented in the BIEMS software package \citep{Mulder:2012}. The methodology is computationally very intensive due to the random selection of many minimal training samples and the fact that equality constraints are approximated with approximate equalities with an arbitrarily small evaluation bound. This limits the applicability of the methodology for general testing in applied research. For this reason an alternative, and computationally cheaper method is proposed here.

The proposed Bayes factor builds on the fractional Bayes factor \cite{OHagan:1995,OHagan:1997,Conigliani:2000}. In the fractional Bayes factor default prior specification is implicitly tackled by splitting the data in a fraction, $b$, that is used for specifying a default (fractional) prior while the remaining fraction, $1-b$, is used for marginal likelihood computation \citep{Gilks:1995}. By setting $b$ to be minimal, maximal information from the data is used for model selection \citep{Berger:1995}.

The original fractional Bayes factor however is not designed for evaluating models with order constraints on the parameters of interest; it was particularly developed for evaluating precise (equality constrained) models. Social science researchers however often formulate their expectations using order constraints (e.g., `group 1 is expected to score higher on average than group 2, and group 2 is expected to score higher on average than group 3' \citep{Klugkist:2005}, or `the effect of the first predictor on the outcome variable is expected to be stronger than the effects of the other predictors' \citep{Braeken:2015}). The problem of the original fractional Bayes factor for order-constrained model selection is that it does not always properly incorporate the complexity of an order-constrained model. To allow for order-constrained model selection using fractional Bayes methodology, the underlying default prior it is centered on the boundary of the constrained space. Given the sets of constraints in \eqref{Mt} this implies that the prior location $\bm\theta_0$ satisfies, $\textbf{R}_t\bm\theta_0=\textbf{r}_t$, where $\textbf{R}_t'=[\textbf{R}_{t,E}'~\textbf{R}_{t,O}']$ and $\textbf{r}_t'=[\textbf{r}_{t,E}'~\textbf{r}_{t,O}']$. Under this adjusted fractional prior, the prior probability of the order constraints, a key quantity in the Bayes factor which quantifies the size of relative complexity of an order-constrained model, would be equal for models with opposite constraints. For example, for the simple test of $H_1:\beta\le0$ versus $H_2:\beta>0$, the prior probabilities of $\beta<0$ or $\beta>0$ will be equal to $\frac{1}{2}$ because the adjusted prior will be centered at 0 under the encompassing prior (while still containing minimal information due to the fraction of the data), thereby appropriately capturing the relative size of the order constrained subspaces. This is not achieved under the default prior in the original fractional Bayes factor which is centered around the likelihood. For this simple test, the proposed default Bayes factor only depends on the posterior probabilities that the constraints hold under a larger unconstrained model \citep[see also][for a related discussion on fractional Bayes factors for one-sided testing]{Lindley:1995}.

The original fractional Bayes factor test is based on a single fraction, typically denoted by $b$. This may result in inconsistent model comparison in case of unbalanced data from different populations for univariate regression models \citep{DeSantis:2001,Hoijtink:2018}. To avoid this inconsistent behavior in the current model selection, the proposed default Bayes factor makes use of group (or population) specific fractions, $b_j$ for group $j$, for $j=1,\ldots,J$, in order to control the amount of information that is used from the data of the different populations for default prior specification. This will allow us to tune the prior to contain minimal information, so that maximal information is available for model selection. The resulting minimally informative default prior has a matrix Cauchy prior distribution. This distribution has not been reported a lot in the literature. Some of its properties have been presented by \cite{BandekarNagar:2003}, for example. This paper presents a new natural application of this distribution for Bayesian model selection problems under the multivariate normal linear model. It will be shown that the proposed default Bayes factor is consistent for the model selection problem in \eqref{Mt} based on this minimally informative default prior.

Another key property of the proposed Bayes factor is that it is relatively fast to compute for the model selection problem in \eqref{Mt}. This is due to the analytic expression of the default Bayes factor for specific cases, and due to the fast Monte Carlo estimate for the general case. The posterior and prior probabilities that the order constraints hold can be computed using available cdf's functions of the multivariate normal and multivariate Student $t$ distribution, and therefore estimates based on the proportions of posterior and prior draws satisfying the constraints, which can be inefficient \citep{Mulder:2012}, can be avoided. Bayesian updating of the evidence in light of new data can also be done relatively fast without needing to store the complete data matrices.

Finally the proposed selection criterion can be computed relatively straightforwardly in the presence of missing data that are missing at random. Although missing data is ubiquitous in statistical practice, it has received surprisingly little attention in model selection problems. When the data contain missing observations, it is widely known that list-wise deletion results in a loss of information and possible bias of the estimates. This can be avoided via multiple imputation mechanisms \citep{Rubin:1987,Rubin:1996}. \cite{Hoijtink:2018b} showed that multiple imputation methods can also be used in certain Bayesian model selection selection problems to avoid bias of the evidence between hypotheses. In this paper we extend this result by showing how the proposed default Bayes factor results in unbiased model selection in the presence of missing data. As imputation only needs to be done under the unconstrained model, the proposed multiple imputation scheme is relatively fast, and thus, multiple imputation under all the competing models (as should be done when using the BIC for example) can be avoided.


This paper is organized as follows. In Section 2 a general formulation of the model is given. Section 3 presents the default Bayes factor for a constrained model selection problem of the form \eqref{Mt}. It is explained how to compute the default Bayes factor, analytic expressions are presented for special cases, consistency is proven as the sample size grows, and the computation is explained in the presence of missing observations. In Section 4 the method is applied to an empirical model selection problem. We end the paper with a short discussion.

\section{Multivariate normal linear model}
The multivariate normal linear model can be written as
\[
\textbf{Y} = \textbf{X}\bm\Theta + \textbf{E}
\]
with $N \times P$ matrix $\textbf{Y}=[\textbf{y}_1\cdots\textbf{y}_N]'$ where $\textbf{y}_i$ is the $i$-th observation of the $P$ dependent observations, $N \times P$ matrix $\textbf{E}=[\textbf{e}_1\cdots\textbf{e}_N]'$ where $\textbf{e}_i\sim N(\textbf{0},\bm\Sigma)$ is the $i$-th $P$ dimensional error vector with unstructured $P\times P$ covariance matrix $\bm\Sigma$, $N \times K$ matrix $\textbf{X}=[\textbf{x}_1\cdots\textbf{x}_N]'$ where $\textbf{x}_i'=(\textbf{d}_i',\textbf{w}_i')$ is the $i$-th observation of the independent variables out of which the first $J$ are dummy variables that indicate group membership, i.e., $d_{ij}=1$ if the $i$-th observation belongs to group $j$ and zero elsewhere, and the remaining $L$ elements are the predictor variables, and matrix of (adjusted) group means and regression coefficients,
\[
\bm{\Theta}=\left[
\begin{array}{ccc}
\mu_{11}&\cdots &\mu_{1P}\\
\vdots & \ddots & \vdots\\
\mu_{J1}&\cdots &\mu_{JP}\\
\beta_{11}&\cdots &\beta_{1P}\\
\vdots & \ddots & \vdots\\
\beta_{L1}&\cdots &\beta_{LP}
\end{array}
\right].
\]
Thus, the $i$-th dependent observation is distributed as follows 
\begin{equation}
\textbf{y}_i\sim N(\bm{\Theta}'\textbf{x}_i,\bm\Sigma).
\label{model1}
\end{equation}

Subsequently, the interest is in a set of $T$ models with competing linear equality and order constraints in \eqref{Mt} on the (adjusted) means and regression coefficients in $\bm\theta=\text{vec}(\bm{\Theta})$. Many common testing problems are special cases of this, such as univariate ($P=1$) or multivariate ($P>1$) linear regression when testing the regression coefficients or interaction effects, univariate or multivariate analysis of variance ($L=0$) or covariance ($L>0$) when testing the (adjusted) group means, or repeated measures models with an unrestricted covariance matrix ($P>2$) when testing the repeated measures means.

\section{A Bayes factor for default model selection}
Following \cite{OHagan:1995}, the marginal likelihood under the original fractional Bayes factor of a constrained model $M_t$ of the form \eqref{Mt} under \eqref{model1} is defined by 
\begin{eqnarray}
\label{marglikeMt} p_t(\textbf{Y},b) &=& \frac{\int_{\bm\Sigma}\int_{\bm{\Theta}\in\mathcal{M}_t} p_t(\textbf{Y}|\textbf{X},\bm{\Theta},\bm\Sigma)\pi_t^N(\bm{\Theta},\bm\Sigma)d\bm{\Theta}d\bm\Sigma}
{\int_{\bm\Sigma}\int_{\bm{\Theta}\in\mathcal{M}_t} p_t(\textbf{Y}|\textbf{X},\bm{\Theta},\bm\Sigma)^b\pi_t^N(\bm{\Theta},\bm\Sigma)d\bm{\Theta}d\bm\Sigma}\\
\label{marglikeMt2}&=& \frac{\int_{\bm\Sigma}\int_{\bm{\Theta}\in\mathcal{M}_t} p(\textbf{Y}|\textbf{X},\bm{\Theta},\bm\Sigma)|\bm\Sigma|^{-\frac{P+1}{2}}d\bm{\Theta}d\bm\Sigma}
{\int_{\bm\Sigma}\int_{\bm{\Theta}\in\mathcal{M}_t} p(\textbf{Y}|\textbf{X},\bm{\Theta},\bm\Sigma)^b|\bm\Sigma|^{-\frac{P+1}{2}}d\bm{\Theta}d\bm\Sigma}
\end{eqnarray}
where $\mathcal{M}_t=\{\bm\theta| \textbf{R}_{t,E}\bm\theta=\textbf{r}_{t,E},~\textbf{R}_{t,O}\bm\theta>\textbf{r}_{t,O} \}$ denotes the parameter subspace that satisfies the constraints under model $M_t$, $p_t(\textbf{Y}|\textbf{X},\bm{\Theta},\bm\Sigma)=p(\textbf{Y}|\textbf{X},\bm{\Theta},\bm\Sigma)I(\bm\Theta\in\mathcal{M}_t)$ denotes the likelihood of the data under model $M_t$, where $p(\textbf{Y}|\textbf{X},\bm{\Theta},\bm\Sigma)$ denotes the likelihood under an unconstrained model, the improper noninformative independence Jeffreys prior is used, i.e., $\pi^N_t(\bm{\Theta},\bm\Sigma)\propto |\bm\Sigma|^{-\frac{P+1}{2}}$, and where
the fraction, $0<b<1$, which controls the amount of information in the data that is used for prior specification, while the remaining fraction, $1-b$, is used for model selection. Note that the likelihood under $M_t$ can be replaced by the unconstrained likelihood in \eqref{marglikeMt2} because the integral is computed over the constrained subspace $\mathcal{M}_t$.

The original fractional Bayes factor is not suitable for order-constrained model selection \citep{Mulder:2014b}. This can be seen as follows. When model $M_t$ only contains order constraints and we consider data that (strongly) supports these constraints, the mass underneath the likelihood and the fraction of the likelihood in \eqref{marglikeMt2} are almost completely located in the order-constrained subspace. Therefore the marginal likelihood will be virtually the same as an unconstrained model without order constraints. The Bayes factor will there for not function as an ``Occam's razor'' where the simpler order-constrained model is preferred over the more complex unconstrained model in the case of an almost equal fit. To correct for this, the integrand in the numerator in \eqref{marglikeMt2} will be integrated over an adjusted parameter space, $\mathcal{M}_t^*=\{\bm\theta| \textbf{R}_{t,E}(\bm\theta-\hat{\bm\theta})=\textbf{0},~\textbf{R}_{t,O}(\bm\theta-\hat{\bm\theta})>\textbf{0}\}$.\footnote{The adjusted parameter space can be seen as a movement of the original constrained parameter space
$\mathcal{M}_t^*=\{\bm\theta| \textbf{R}_{t,E}(\bm\theta+\hat{\bm\theta}-\bm\theta_0)=\textbf{r}_{t,E},~\textbf{R}_{t,O}(\bm\theta+\hat{\bm\theta}-\bm\theta_0)>\textbf{r}_{t,O}\}$, where $\bm\theta_0$ satisfies $[\textbf{R}_{t,E}'~\textbf{R}_{t,O}']'\bm\theta_0=[\textbf{r}_{t,E}'~\textbf{r}_{t,O}']'$.} This (implicitly) results in a default prior that is centered on the boundary of the constrained space having prior uncertainty based on a fraction of the information in the prior.

Furthermore, to properly control the amount of information that is used for default prior specification and to avoid inconsistent behavior, different fractions are needed for observations coming different groups \citep{DeSantis:2001,Hoijtink:2018b}. For this reason, group specific fractions will be used for the likelihood that is raised to $b$ in the numerator in \eqref{marglikeMt}. The generalized fraction of the likelihood will be defined by
\[
p(\textbf{Y}|\textbf{X},\bm{\Theta},\bm\Sigma)^{\textbf{b}}\equiv
\prod_{i=1}^N
p(\textbf{y}_i|\textbf{x}_i,\bm{\Theta},\bm\Sigma)^{b_i}.
\]
We come back to the choice of the fractions later in this paper.

Consequently, the proposed default marginal likelihood can therefore be expressed as
\begin{eqnarray}
\label{marglikeMt3} p^*_t(\textbf{Y},\textbf{b}) &=& \frac{\int_{\bm\Sigma}\int_{\bm{\Theta}\in\mathcal{M}_t} p(\textbf{Y}|\textbf{X},\bm{\Theta},\bm\Sigma)|\bm\Sigma|^{-\frac{P+1}{2}}d\bm{\Theta}d\bm\Sigma}
{\int_{\bm\Sigma}\int_{\bm{\Theta}\in\mathcal{M}^*_t} p(\textbf{Y}|\textbf{X},\bm{\Theta},\bm\Sigma)^{\textbf{b}}|\bm\Sigma|^{-\frac{P+1}{2}}d\bm{\Theta}d\bm\Sigma}.
\end{eqnarray}
The default Bayes factor of a constrained model of the form $M_t$ against an unconstrained model $M_u$ can then be expressed as a multivariate Savage-Dickey density ratio \citep{Dickey:1971} multiplied with a ratio of posterior and prior probabilities that the order constraints hold conditional on that the equality constraints hold.

\begin{lemma}
The default Bayes factor for a constrained model $M_t$ of the form \eqref{Mt} against an unconstrained alternative model $M_u$ based on the marginal likelihood in \eqref{marglikeMt} can be expressed as
\begin{eqnarray}
B_{tu} &=& \frac{f^E_t(\textbf{Y},\textbf{X})}{c^E_t(\textbf{Y}_{\textbf{b}},\textbf{X}_{\textbf{b}})}\times
\frac{f^{O}_t(\textbf{Y},\textbf{X})}{c^{O}_t(\textbf{Y}_{\textbf{b}},\textbf{X}_{\textbf{b}})},
\label{Btu}
\end{eqnarray}
where
\begin{eqnarray*}
f^E_t(\textbf{Y},\textbf{X}) &=&
\pi_u(\textbf{R}_{t,E}\bm\theta=\textbf{r}_{t,E}|\textbf{Y},\textbf{X})\\
c^E_t(\textbf{Y},\textbf{X},{\textbf{b}}) &=& 
\pi_u^*(\textbf{R}_{t,E}\bm\theta=\textbf{r}_{t,E}|\textbf{Y},\textbf{X},{\textbf{b}})\\
f^{O}_t(\textbf{Y},\textbf{X}) &=&
\text{Pr}_u(\textbf{R}_{t,O}\bm\theta>\textbf{r}_{t,O}|\textbf{R}_{t,E}\bm\theta=\textbf{r}_{t,E},\textbf{Y},\textbf{X})\\
c^{O}_t(\textbf{Y},\textbf{X},{\textbf{b}}) &=&
\text{Pr}^*_u(\textbf{R}_{t,O}\bm\theta>\textbf{r}_{t,O}|\textbf{R}_{t,E}\bm\theta=\textbf{r}_{t,E},\textbf{Y},\textbf{X},{\textbf{b}}),
\end{eqnarray*}
where the marginal unconstrained posterior and default prior for $\bm\Theta$ under $M_u$ follow matrix $t$ distributions given by
\begin{eqnarray}
\label{uncpost}\pi_u(\bm{\Theta}|\textbf{Y},\textbf{X}) &=& \mathcal{T}_{K\times P}(
\hat{\bm{\Theta}}, ({\textbf{X}}'{\textbf{X}})^{-1}, {\textbf{S}},N-K-P+1)\\
\label{fracprior}\pi^*_u(\bm{\Theta}|\textbf{Y},\textbf{X},{\textbf{b}}) &=& \mathcal{T}_{K\times P}(
\bm{\Theta}_0,({\textbf{X}}'_{\textbf{b}}{\textbf{X}}_{\textbf{b}})^{-1},{\textbf{S}}_{\textbf{b}},\sum_{i=1}^N b_i-K-P+1),
\end{eqnarray}
from which the conditional and marginal distributions used for computing $f^E_t$, $f^{O}_t$, $c^E_t$, and $c^{O}_t$ naturally follow, and where $\bm{\theta}_0=\text{vec}(\bm{\Theta}_0)$ satisfies $[\textbf{R}_{t,E}'~\textbf{R}'_{t,O}]'\bm\theta_0=[\textbf{r}_{t,E}'~\textbf{r}_{t,O}']'$, the LS estimate equals $\hat{\bm{\Theta}}=(\textbf{X}'{\textbf{X}})^{-1}{\textbf{X}}'{\textbf{Y}}$, the sums of square matrix in the posterior equals ${\textbf{S}}=({\textbf{Y}} - {\textbf{X}}\hat{{\bm{\Theta}}})'({\textbf{Y}} - {\textbf{X}}\hat{{\bm{\Theta}}})$, and the sums of square matrix in the default prior equals ${\textbf{S}}_{\textbf{b}}=({\textbf{Y}}_{\textbf{b}} - {\textbf{X}}_{\textbf{b}}\hat{{\bm{\Theta}}}_{\textbf{b}})'({\textbf{Y}}_{\textbf{b}} - {\textbf{X}}_{\textbf{b}}\hat{{\bm{\Theta}}}_{\textbf{b}})$, with $\hat{{\bm{\Theta}}}_{\textbf{b}}=({\textbf{X}}'_{\textbf{b}}{\textbf{X}}_{\textbf{b}})^{-1}{\textbf{X}}_{\textbf{b}}'{\textbf{Y}}_{\textbf{b}}$, where $\textbf{Y}_{\textbf{b}}$ and $\textbf{X}_{\textbf{b}}$ are the stacked matrices of $\textbf{y}_{i,b_i}'$ and $\textbf{x}_{i,b_i}'$, with ${\textbf{y}}_{i,b_i}=\sqrt{b_i}\textbf{y}_i$ and ${\textbf{x}}_{i,b_i}=\sqrt{b_i}\textbf{x}_i$.
\end{lemma}
A proof is given in Appendix A. The posterior quantities in the numerators in \eqref{Btu} are denoted with a ``$f$'' because they can be interpreted as measures of relative fit of the constrained model $M_t$ relative to the unconstrained model. The prior quantities in the denominators in \eqref{Btu} are measures of relative complexity of $M_t$ and therefore denoted with a ``$c$''. Furthermore the superscript of these symbols denote which part of the constraints of model $M_t$ it evaluates (either the equality constraints ``$E$'' or the order constraints conditional that the equality constraints hold ``$O$''). Because the location hyperparameter in the unconstrained default prior $\pi^*_u$ satisfies $[\textbf{R}_{t,E}'~\textbf{R}_{t,O}']'\bm\theta_0=[\textbf{r}_{t,E}'~\textbf{r}_{t,O}']'$, we say that the prior is centered on the boundary of the constrained space of $M_t$.

\subsection{Bayes factor computation}
Unlike the matrix normal distribution, the equivalent marginal posterior of the vectorization $\bm\theta$ does note follow a multivariate Student $t$ distribution; only the marginal distributions of the separate columns or rows of $\bm{\Theta}$ have multivariate $t$ distributions \citep[][p. 443]{BoxTiao}. Consequently, a linear combination of the elements in $\bm\theta$, say, $\bm\zeta_{t,E}=\textbf{R}_{t,E}\bm\theta$, does not have a multivariate student $t$ distribution or other known distributional form for a coefficient matrix $\textbf{R}_{t,E}$ in general. Therefore, the posterior density in the numerator in the first term in \eqref{Btu} does not have an analytic form. A Monte Carlo estimate can be obtained relatively easy however. First we define the one-to-one transformation, $\bm\zeta_{t}=(\bm\zeta_{t,E}',\bm\zeta_{t,O}')'=\textbf{H}_t\bm\theta$, where $\textbf{H}_t=[\textbf{R}_{t,E}' ~\textbf{D}_t' ]'$, with $\textbf{D}_t$ being a $(PK-r_{t,E})\times PK$
matrix such that $\bm\zeta_{t,O}$ contains the parameters in $\bm\theta$ that not constrained with equalities. Conditionally on $\bm\Sigma$, the transformed parameters, $\bm\zeta_{t}$, have a multivariate normal conditional posterior, $\mathcal{N}(\bm\mu_{\bm\zeta_{t}},\bm\Psi_{\bm\zeta_{t}})$, with $\bm\mu_{\bm\zeta_{t}}=\textbf{H}_t\hat{\bm\theta}$ and $\bm\Psi_{\bm\zeta_{t}}=\textbf{H}_t[\bm\Sigma\otimes ({\textbf{X}}'{\textbf{X}})^{-1}]\textbf{H}_t'$. Then,
\begin{eqnarray}
\label{densMC} \pi_u(\textbf{R}_{t,E}\bm\theta=\textbf{r}_{t,E}|\textbf{Y},\textbf{X}) &=&
\int\pi_u(\bm\zeta_{t,E}=\textbf{r}_{t,E},\bm\zeta_{t,O}|\textbf{Y},\textbf{X}) d\bm\zeta_{t,O}\\
\nonumber&=&
\iint \pi_u(\bm\zeta_{t,E}=\textbf{r}_{t,E},\bm\zeta_{t,O}|\textbf{Y},\textbf{X},\bm\Sigma)
 \pi_u(\bm\Sigma|\textbf{Y},\textbf{X})d\bm\Sigma d\bm\zeta_{t,O}\\
\nonumber&\approx &S^{-1}\sum_{s=1}^S \mathcal{N}_{PK}(\textbf{r}_{t,E};\textbf{R}_{t,E}\hat{\bm\theta},\textbf{R}_{t,E}[\bm\Sigma^{(s)}\otimes ({\textbf{X}}'{\textbf{X}})^{-1}]\textbf{R}_{t,E}'),
\end{eqnarray}
where $\bm\Sigma^{(s)}\sim\mathcal{IW}(N-K,\textbf{S})$, for $s=1,\ldots,S$, and $\mathcal{N}_{PK}(\textbf{r};\bm\mu,\bm\Psi)$ denotes a $PK$-variate normal density with mean vector $\bm\mu$ and covariance matrix $\bm\Psi$ evaluated at $\textbf{r}$, which has an analytic expression (in {\tt R}, for instance, it can be computed using the {\tt dmvnorm} function from the {\tt mvtnorm} package). This Monte Carlo estimate can be obtained via parallelized computation, and it is therefore computationally cheap.

The conditional posterior probability in the numerator in the second term can be obtained in a similar manner. First note that the order constraints under $M_t$ for the transformed parameter vector are equivalent to $\tilde{\textbf{R}}_{t,O}\bm\zeta_{t,O}>\textbf{r}_{t,O}$, where $\tilde{\textbf{R}}_{t,O}$ consists of the columns corresponding to the parameters of $\bm\theta$ that are not constrained with equalities.
%
Furthermore, a property of the multivariate normal distribution is that the conditional posterior of $\bm\zeta_{t,O}$ given $\bm\zeta_{t,E}=\textbf{r}_{t,E}$ has a multivariate normal distribution with mean $\bm\mu_{\bm\zeta_{t},O|E}=\bm\mu_{\bm\zeta_{t},O}$+$\bm\Psi_{\bm\zeta_{t},OE}\bm\Psi_{\bm\zeta_{t},EE}^{-1}(\textbf{r}_{t,E}-\bm\mu_{\bm\zeta_{t},E})$ and covariance matrix $\bm\Psi_{\bm\zeta_{t},O|E}=\bm\Psi_{\bm\zeta_{t},OO}-\bm\Psi_{\bm\zeta_{t},OE}\bm\Psi_{\bm\zeta_{t},EE}^{-1}\bm\Psi_{\bm\zeta_{t},EO}$, where the indices $E$ and $O$ refer to the appropriate parts of the mean vector and covariance matrix of $\bm\zeta_t$. Now when $\tilde{\textbf{R}}_{t,O}$ is of full row rank, which is generally the case\footnote{An example of a set of order constraints that corresponds to a matrix $[\textbf{R}_{t,O}|\textbf{r}_{t,O}]$ that is not of full row rank is when a subset of parameters is expected to be larger/smaller than another set of parameters, e.g., $(\beta_1,\beta_2)>(\beta_3,\beta_4)$.}, a transformed parameter can be defined, $\bm\eta_{t,O}=\tilde{\textbf{R}}_{t,O}\bm\zeta_{t,O}$, which has a multivariate normal distribution with mean vector $\bm\mu_{\bm\eta_{t},O}=\tilde{\textbf{R}}_{t,O}\bm\mu_{\bm\zeta_{t},O|E}$ and covariance matrix $\bm\Psi_{\bm\eta_{t},O}=\tilde{\textbf{R}}_{t,O}\bm\Psi_{\bm\zeta_{t},O|E}\tilde{\textbf{R}}_{t,O}'$. The posterior probability in the numerator in the second term in \eqref{Btu} can then be computed via a Monte Carlo estimate, similar as in \eqref{densMC},
\[
\text{Pr}_u(\textbf{R}_{t,O}\bm\theta>\textbf{r}_{t,O}|\textbf{R}_{t,E}\bm\theta=\textbf{r}_{t,E},\textbf{Y},\textbf{X}) \approx S^{-1}\sum_{s=1}^S \Phi_{\mathcal{N}}(-\textbf{r}_{t,O};-\bm\mu_{\bm\eta_{t},O},\bm\Psi_{\bm\eta_{t},O}^{(s)}),
\]
where $\bm\Psi_{\bm\eta_{t},O}^{(s)}$ is computed using $\bm\Psi_{\bm\zeta_{t}}^{(s)}=\textbf{H}_t[\bm\Sigma^{(s)}\otimes ({\textbf{X}}'{\textbf{X}})^{-1}]\textbf{H}_t'$, with $\bm\Sigma^{(s)}\sim\mathcal{IW}(N-K,\textbf{S})$, for $s=1,\ldots,S$, and $\Phi_{\mathcal{N}}$ denotes the multivariate normal cdf. Note that the cdf can be computed using standard statistical software (e.g., using the {\tt pmvnorm} function from the {\tt mvtnorm} package in {\tt R}). When $\textbf{R}_{t,O}$ is not of full row rank, the above method cannot be used. To get the conditional posterior probability, a numerical sampling estimate can be used via the {\tt R}-function {\tt bain} in the {\tt bain} package \citep{Gu:2019}.

Note that if the sample size is sufficiently large, the matrix $t$ can be well approximated using a matrix normal distribution, $\mathcal{N}_{K\times P}(\hat{\bm{\Theta}},(\textbf{X}'\textbf{X})^{-1},(N-K-P+1)^{-1}\textbf{S})$ \citep[][p. 447]{BoxTiao}. In this case no Monte Carlo estimate would be needed as the posterior density and the posterior probability could directly be computed using the approximated matrix (or multivariate) normal distribution of $\bm{\Theta}$ (or $\bm\theta$).

\subsection{Minimally informative default prior}
The group specific fractions will be chosen such that the information in the default prior is minimal, so that maximal information in the data is used for model selection. Following \cite{Berger:1995}, alternative choices for the fractions, such as $b=\log n$ or $b=\sqrt{n}$ as suggested by \citep{OHagan:1995}, would not be recommendable as the amount of information in the prior would then diverge as the sample size grows.

The group dependent fractions are chosen such that from each group the same amount of information, say, the information in $m$ independent obervations, is taken. This can be achieved by setting $b_{i}=\frac{m}{n_j}$ if the $i$-th observation belongs to the $j$-th group, i.e., $d_{ij}=1$, where $n_j$ is the sample size of group $j$. Note that in the original fractional Bayes factor with a single fraction for one group, the minimal fraction would be $b=\frac{m}{n}$ where $m$ is the minimal sample size to get a finite marginal likelihood. Given the matrix $t$ distribution of the default prior for $\bm{\Theta}$ in \eqref{fracprior}, a minimally informative prior is obtained when the prior degrees of freedom of this distribution is equal to one, i.e, $\sum_{i=1}^n b_i-K-P+1=1$. This is achieved by setting $m=\frac{P+K}{J}$. To keep the notation simple, $\textbf{b}$ will denote the minimal fraction throughout the remainder of the paper. The marginal fractional prior for $\bm{\Theta}$ in \eqref{fracprior} then has a matrix Cauchy distribution,
\begin{equation}
\label{fracprior2}\pi_u^*(\bm{\Theta}|\textbf{Y},\textbf{X},\textbf{b})=\mathcal{C}_{K\times P}(
\bm{\Theta}_0, {\textbf{S}}_{\textbf{b}}, ({\textbf{X}}'_{\textbf{b}}{\textbf{X}_{\textbf{b}}})^{-1}).
\end{equation}
Note that the matrix Cauchy distribution has not yet received a lot of attention in the literature. Here we see an interesting application where it naturally appears as a minimally informative default prior for the coefficients matrix $\bm\Theta$.

The prior density in the denominator in the first term in \eqref{Btu} and the prior probability in the denominator in the second term in \eqref{Btu} based on the above matrix Cauchy prior can be computed using the same Monte Carlo estimate as was shown for their respective posterior counterparts. Note that if the order constraints under $M_t$ are solely specified between parameters in the same column or same row of $\bm{\Theta}$, the prior probability can also be computed using a multivariate normal distribution with the same location and covariance structure. This is because the probability of a set of order constraints is invariant to the exact distributional form as long as the mean lies on the boundary, and an elliptical distribution is used with the same covariance structure. Finally note that because the amount of prior information is kept minimal regardless of the sample size, the default prior cannot be approximated with a matrix normal distribution for larger sample, unlike the posterior.

\subsection{Analytic expression for special cases}
The marginal distribution of a column (or row\footnote{Note that when $\bm{\Theta}\sim \mathcal{T}_{K,P}(\nu,\textbf{M},\bm\Psi,\bm\Phi)$, then $\bm{\Theta}'\sim \mathcal{T}_{P,K}(\nu,\textbf{M}',\bm\Phi,\bm\Psi)$ \citep[][p. 442]{BoxTiao}}) of a matrix random variable with a matrix Student $t$ distribution has a multivariate Student $t$ distribution \citep[][p. 442-443]{BoxTiao}. This implies that the unconstrained marginal prior and posterior of the $p$-th column of $\bm{\Theta}$, denoted by $\bm{\theta}_p$, are distributed as
\begin{eqnarray*}
\pi_u(\bm{\theta}_{p}|\textbf{Y},\textbf{X}) &=& \mathcal{T}_K(\hat{\bm{\theta}}_{p},(N-K-P+1)^{-1}s_{pp}(\textbf{X}'\textbf{X})^{-1},N-K-P+1)\\
\pi^*_u(\bm{\theta}_{p}|\textbf{Y},\textbf{X},\textbf{b}) &=& \mathcal{C}_K(\bm{\theta}_{0,p},s_{\textbf{b},pp}(\textbf{X}'_{\textbf{b}}\textbf{X}_{\textbf{b}})^{-1}),
\end{eqnarray*}
where $s_{\textbf{b},pp}$ and $s_{pp}$ denote the $(p,p)$-th element of $\textbf{S}_{\textbf{b}}$ and $\textbf{S}$, respectively. Thus, using standard calculus, it can be shown that for a constrained model with only constraints on the elements in column $p$, i.e., $M_t:\textbf{R}_{t,E}\bm\theta_p=\textbf{r}_{t,E}~\&~\textbf{R}_{t,O}\bm\theta_p>\textbf{r}_{t,O}$, the posterior and prior quantities in \eqref{Btu} are equal to
\begin{eqnarray*}
f^E_t(\textbf{Y},\textbf{X}) &=&
\mathcal{T}(\textbf{r}_{t,E};\textbf{R}_{t,E}\hat{\bm{\theta}}_{p},(N-K-P+1)^{-1}s_{pp}\textbf{R}_{t,E}(\textbf{X}'\textbf{X})^{-1}\textbf{R}_{t,E}',\\
&&N-K-P+1)
\\
c^E_t(\textbf{Y},\textbf{X},{\textbf{b}}) &=& \mathcal{C}(\textbf{r}_{t,E};\textbf{R}_{t,E}\bm{\theta}_{0,p},s_{\textbf{b},pp}\textbf{R}_{t,E}(\textbf{X}'_{\textbf{b}}\textbf{X}_{\textbf{b}})^{-1}\textbf{R}_{t,E}')\\
f^{O}_t(\textbf{Y},\textbf{X}) &=& 
\Psi_{\mathcal{T}}(-\textbf{r}_{t,O};-\bm\mu_{\bm\eta_{t},O}^f,\bm\Psi^f_{\bm\eta_{t},O},N-K-P+1)\\
c^{O}_t(\textbf{Y},\textbf{X},{\textbf{b}}) &=& 
\Psi_{\mathcal{C}}(\textbf{0};\textbf{0},\bm\Psi^c_{\bm\eta_{t},O})
\end{eqnarray*}
where $\Psi_{\mathcal{T}}(\textbf{x};\bm\mu,\bm\Psi,\nu)$ is the cdf of a multivariate Student $t$ distribution at $\textbf{x}$ with location $\bm\mu$, scale matrix $\bm\Psi$, and $\nu$ degrees of freedom, $\Psi_{\mathcal{C}}(\textbf{x};\bm\mu,\bm\Psi)$ is the cdf of a multivariate Cauchy distribution at $\textbf{x}$ with location $\bm\mu$ and scale matrix $\bm\Psi$, and  
\begin{eqnarray*}
\bm\mu_{\bm\eta_{t},O}^f &=& \textbf{R}_{t,O}\hat{\bm\theta}_p+\textbf{R}_{t,O}(\textbf{X}'\textbf{X})^{-1}\textbf{R}_{t,E}'(\textbf{R}_{t,E}(\textbf{X}'\textbf{X})^{-1}\textbf{R}_{t,E}')^{-1}(\textbf{r}_{t,E}-\textbf{R}_{t,E}\hat{\bm\theta}_p)\\
\bm\Psi_{\bm\eta_{t},O}^f&=&\frac{s_{pp}+(\textbf{r}_{t,E}-\textbf{R}_{t,E}\hat{\bm\theta}_p)'(\textbf{R}_{t,E}(\textbf{X}'\textbf{X})^{-1}\textbf{R}_{t,E}')^{-1}(\textbf{r}_{t,E}-\textbf{R}_{t,E}\hat{\bm\theta}_p)}{N-K-P+1+r_{t,E}}\\
&&\textbf{R}_{t,O}(\textbf{X}'\textbf{X})^{-1}\left(\textbf{I}_{K-r_{t,E}}-\textbf{R}_{t,E}'(\textbf{R}_{t,E}(\textbf{X}'\textbf{X})^{-1}\textbf{R}_{t,E}')^{-1}\textbf{R}_{t,E}(\textbf{X}'\textbf{X})^{-1}\right)\textbf{R}_{t,O}'\\
\bm\Psi_{\bm\eta_{t},O}^c&=&\tfrac{s_{\textbf{b},pp}}{1+r_{t,E}}\textbf{R}_{t,O}(\textbf{X}'_{\textbf{b}}\textbf{X}_{\textbf{b}})^{-1}\left(\textbf{I}_{K-r_{t,E}}-(\textbf{X}'_{\textbf{b}}\textbf{X}_{\textbf{b}})^{-1}\textbf{R}_{t,E}'(\textbf{R}_{t,E}(\textbf{X}'_{\textbf{b}}\textbf{X}_{\textbf{b}})^{-1}\textbf{R}_{t,E}')^{-1}\right.\\
&&\left.\textbf{R}_{t,E}(\textbf{X}_{\textbf{b}}'\textbf{X}_{\textbf{b}})^{-1}\right)\textbf{R}_{t,O}',
\end{eqnarray*}
where $r_{t,E}$ is the number of rows of $\textbf{R}_{t,E}$.
Note that the cdf's can be computed using standard functions in statistical software (e.g., using {\tt pmvt} in the {\tt mvtnorm}-package \citep{Genz:2016})

Hence, the proposed default Bayes factor has an analytic expression for univariate testing problems (e.g., AN(C)OVA or linear regression), for multivariate/univariate $t$ tests, or in other testing problems where the constraints are formulated solely on the elements of one specific column or row of $\bm{\Theta}$.

\subsection{Consistency}
A Bayes factor is called consistent if the evidence goes to infinity for the true constrained model against the alternative models as the sample goes to infinity. Consistency is therefore a fundamental property that a model selection criterion should have because it ensures that the true model will always be selected as long as the sample is large enough.

\begin{lemma}
Given a set of competing nonnested multivariate normal linear models with equality and order constraints on the location parameters $\bm\theta$ of the form \eqref{Mt}, the default Bayes factor defined in \eqref{Btu} with group specific minimal fractions is consistent.
\end{lemma}
Here a sketch of the proof is given. First note that the unconstrained posterior density in the numerator in the first term in \eqref{Btu} goes to infinity if the equality constraints hold, and to zero if they do not hold. Second, the conditional posterior probability in the numerator in the second term goes to 1 if the constraints hold and to 0 if they do not hold. The quantities in the denominators depend on the unconstrained matrix Cauchy prior in \eqref{fracprior2} with scale matrices $(\textbf{X}'_{\textbf{b}}\textbf{X}_{\textbf{b}})^{-1}$ and $\textbf{S}_{\textbf{b}}$. As the sample sizes, $n_j$ for all groups go to infinity, these scale matrices converge to finite scale matrices which depend on the population distributions of the observed variables. For example, the scale matrix $\sum_{j=1}^Jn_j^{-1}\textbf{X}_{j}'\textbf{X}_{j}$ in \eqref{XXb} converges to
\[
\left[
\begin{array}{cccc}
1 & 0 & 0 & E\{\textbf{w}'_1\}\\
0 & \ddots & 0 & \vdots\\
0 & 0 & 1 & E\{\textbf{w}'_J\}\\
E\{\textbf{w}_1\} & \cdots & E\{\textbf{w}_J\} & \sum_j E\{\textbf{w}_j\textbf{w}_j'\} 
\end{array}
\right].
\]
Similar results hold for the matrices $\sum_{j=1}^Jn_j^{-1}\textbf{X}_{j}'\textbf{Y}_{j}$ and $\sum_{j=1}^Jn_j^{-1}\textbf{Y}_{j}'\textbf{Y}_{j}$ in \eqref{Sb} in the limit. The unconstrained default prior distribution therefore converges to some fixed matrix Cauchy distribution. This implies that the value of the prior density in the denominator in the first term in \eqref{Btu} converges to some positive constant as well as the conditional prior probability in the numerator in the second term for all constrained models under consideration. Consequently the Bayes factor $B_{tu}$ for a true constrained model $M_t$ goes to infinity, and $B_{tu}$ converges to zero for an incorrect model. The proposed default Bayes factor is therefore consistent.

\subsection{Sequential Bayesian updating}
Similar as the original fractional Bayes factor, the proposed default Bayes factor is not coherent when sequential updating of the evidence when observing new data $(\textbf{Y}_{new},\textbf{X}_{new})$. This implies that the evidence has to be recomputed when new data are observed because the fractional prior will (slightly) change. This can be done very efficiently as discussed below.

The default fractional prior in \eqref{fracprior} depends on the data via the sufficient statistics,
\begin{eqnarray}
\label{XXb}{\textbf{X}}'_{\textbf{b}}{\textbf{X}_{\textbf{b}}} = \sum_{i=1}^n b_i\textbf{x}_i\textbf{x}_i'=\tfrac{P+K}{J}\sum_{j=1}^J n_j^{-1}\textbf{X}_j'\textbf{X}_j,
\end{eqnarray}
where (with a slight abuse of notation) $\textbf{X}_j$ denotes the stacked matrix of the covariates $\textbf{x}_i'$ for group $j$, and similarly for the fractional sums of squares matrix,
\begin{eqnarray}
\label{Sb}{\textbf{S}}_{\textbf{b}} &=& \textbf{Y}'_{\textbf{b}}\textbf{Y}_{\textbf{b}} - \textbf{Y}'_{\textbf{b}}\textbf{X}_{\textbf{b}}(\textbf{X}_{\textbf{b}}'\textbf{X}_{\textbf{b}})^{-1}
\textbf{X}_{\textbf{b}}'\textbf{Y}_{\textbf{b}}\\
\nonumber &=& \tfrac{P+K}{J}\left(\sum_{j=1}^J n_j^{-1}\textbf{X}_j'\textbf{Y}_j
-(\sum_{j=1}^J n_j^{-1}\textbf{Y}_j'\textbf{X}_j)(\sum_{j=1}^J n_j^{-1}\textbf{X}_j'\textbf{X}_j)^{-1}(\sum_{j=1}^J n_j^{-1}\textbf{X}_j'\textbf{Y}_j)
\right),
\end{eqnarray}
where (with a slight abuse of notation) $\textbf{Y}_j$ denotes the stacked matrix of the outcome variables $\textbf{y}'_i$ for group $j$. Thus, when observing new data matrices, $\textbf{X}_{new,j}$ and $\textbf{Y}_{new,j}$, we only need to update the group specific sufficient statistics, i.e.,
\begin{eqnarray*}
\textbf{X}_j'\textbf{X}_j &\rightarrow &\textbf{X}_j'\textbf{X}_j + \textbf{X}_{new,j}'\textbf{X}_{new,j}\\
\textbf{X}_j'\textbf{Y}_j &\rightarrow &\textbf{X}_j'\textbf{Y}_j + \textbf{X}_{new,j}'\textbf{Y}_{new,j}\\
\textbf{Y}_j'\textbf{Y}_j &\rightarrow &\textbf{Y}_j'\textbf{Y}_j + \textbf{Y}_{new,j}'\textbf{Y}_{new,j},
\end{eqnarray*}
to obtain the updated fractional prior.

The unconstrained posterior depends on the same sufficient statistics for the complete data set, i.e., $\textbf{X}'\textbf{X}$, $\textbf{X}'\textbf{Y}$, and $\textbf{Y}'\textbf{Y}$. These can be updated in a similar manner as above when observing new data.

\subsection{Missing data}
Model selection in the presence of missing observation has received surprisingly little attention in the literature despite the ubiquity of missing data in applied research. The literature on missing data handling \citep[e.g., multiple imputation;][]{Rubin:1987,Rubin:1996} has mainly focused on estimation problems in the presence of missing data. To make using of multiple imputation mechanisms in the presence of model uncertainty, which is the situation we consider here, one possibility would be to compute the marginal likelihoods by constructing complete imputed datasets where the imputation models correspond to the competing models under investigation, possibly with the addition of auxiliary variables. This however will be computationally very expensive as different imputation models would be needed under the different constrained models. A complicating factor when drawing random imputed observations is that the model parameters have to satisfy different sets of order constraints. In this section we show how to avoid these difficulties by computing the proposed default Bayes factors between the $T$ constrained models using a single unconstrained imputation model. This result extends the work of Hoijtink et al. (2018) who focused on an approximated Bayes factor based on normal approximations.

Let us denote the observed data matrices by $\textbf{Y}^{o}$ and $\textbf{X}^{o}$, and the data matrices with the missing observations by $\textbf{Y}^{m}$ and $\textbf{Y}^{m}$. Because of the missing observations, no analytic expressions are available for the sufficient statistics to compute the Bayes factor in \eqref{Btu}. This Bayes factor however is only a function of four quantities, $f^E_t$, $f^O_t$, $c^E_t$, and $c^O_t$, which are integrals of the unconstrained posterior \eqref{uncpost} or unconstrained default prior \eqref{fracprior} over the equality or order constrained subspace of model $M_t$. Thus, if we can obtain unbiased estimates of these unconstrained distributions, we can obtain unbiased estimates of the four quantities, and consequently, we obtain unbiased estimates of the relative evidence between the hypotheses via the Bayes factor. Thus the model selection problem in the presence of missing data now has become an estimation problem in the present of missing data under the unconstrained model. Now we can utilize the huge body of literature on obtaining unbiased estimates (under an unconstrained model) in the presence of missing data via multiple imputation techniques \citep{Rubin:1987,Rubin:1996}. To get an unbiased estimate of the relative fit measure $f^E_t$, i.e., the fit of the equality constraints of model $M_t$ relative to the unconstrained model, we can compute the arithmetic average of these measures based on many randomly generated complete data sets via multiple imputation under the unconstrained model, 

\begin{eqnarray*}
f^E_t(\textbf{Y}^{o},\textbf{X}^{o}) &=& \pi_u(\textbf{R}_{E}\bm\theta=\textbf{r}_{E}|\textbf{Y}^{o},\textbf{X}^{o})\\
&=& \iint \pi_u(\textbf{R}^E\bm\theta=\textbf{r}^E|\textbf{Y}^{o},\textbf{Y}^{m},\textbf{X}^{o},\textbf{X}^{m}) \pi_u(\textbf{Y}^{m},\textbf{X}^{m}|\textbf{Y}^{o},\textbf{X}^{o}) d\textbf{X}^{m}d\textbf{Y}^{m}\\
&\approx & M^{-1}\sum_{m=1}^M \pi_u(\textbf{R}^E\bm\theta=\textbf{r}^E|\textbf{Y}^{o},\textbf{Y}^{(m)},\textbf{X}^{o},\textbf{X}^{(m)}),\\
&\approx &M^{-1}\sum_{m=1}^M N(\textbf{r}_E;\textbf{R}_E\hat{\bm\theta}^{o,(m)},\textbf{R}_E[\bm\Sigma^{(m)}\otimes ({\textbf{X}^{o,(m)}}'{\textbf{X}^{o,(m)}})^{-1}]\textbf{R}_E')
\end{eqnarray*}
where $\bm\Sigma^{(m)}\sim\mathcal{IW}(N-K,\textbf{S}^{o,(m)})$, with $\textbf{S}^{o,(m)}=({\textbf{Y}^{o,(m)}}-{\bm{\Theta}^{o,(m)}}{\textbf{X}^{o,(m)}})'({\textbf{Y}^{o,(m)}}-{\bm{\Theta}^{o,(m)}}{\textbf{X}^{o,(m)}})$, and ${\bm{\Theta}^{o,(m)}}=({\textbf{X}^{o,(m)}}'{\textbf{X}^{o,(m)}})^{-1}{\textbf{X}^{o,(m)}}{\textbf{Y}^{o,(m)}}$,
$\textbf{Y}^{(m)}$ and $\textbf{X}^{(m)}$ are the $m$-th draws of the data matrices of the dependent variables and predictor variables with missing observations, respectively, sampled from the unconstrained posterior $\pi_u(\textbf{Y}^{m},\textbf{X}^{m}|\textbf{Y}^{o},\textbf{X}^{o})$, and $\textbf{Y}^{o,m}$ and $\textbf{X}^{o,m}$ denote the complete data matrix that combines the observed and missing data matrices. 
Similarly, the measures of relative complexity of the equality constraints of $M_t$ can be obtained via
\begin{eqnarray*}
c^E_t(\textbf{Y}^{o},\textbf{X}^{o},\textbf{b}) &=& \pi^*_u(\textbf{R}_{E}\bm\theta=\textbf{r}_{E}|\textbf{Y}^{o},\textbf{X}^{o},\textbf{b})\\
&=& \iint \pi^*_u(\textbf{R}^E\bm\theta=\textbf{r}^E|\textbf{Y}^{o},\textbf{Y}^{m},\textbf{X}^{o},\textbf{X}^{m},\textbf{b}) \pi_u(\textbf{Y}^{m},\textbf{X}^{m}|\textbf{Y}^{o},\textbf{X}^{o}) d\textbf{X}^{m}d\textbf{Y}^{m}\\
&\approx & M^{-1}\sum_{m=1}^M \pi_u^*(\textbf{R}^E\bm\theta=\textbf{r}^E|\textbf{Y}^{o},\textbf{Y}^{(m)},\textbf{X}^{o},\textbf{X}^{(m)},\textbf{b}),\\
&\approx &M^{-1}\sum_{m=1}^M N(\textbf{r}_E;\textbf{R}_E\bm\theta_0,\textbf{R}_E[\bm\Sigma^{(m)}\otimes (\textbf{X}^{o,(m)'}_{\textbf{b}}{\textbf{X}_{\textbf{b}}^{o,(m)}})^{-1}]\textbf{R}_E'),
\end{eqnarray*}
where $\bm\Sigma^{(m)}\sim\mathcal{IW}(P,\textbf{S}^{o,(m)}_{\textbf{b}})$. Note that the sampling distribution of the missing observations, $\pi_u(\textbf{Y}^{m},\textbf{X}^{m}|\textbf{Y}^{o},\textbf{X}^{o})$, is the same as used for the posterior. Thus there is only a single imputation model to perform the model selection. Finally note that the imputation model may involve additional auxiliary variables which are not included in the analysis model.

\section{Empirical applications}
\subsection{One way ANOVA}
Informative hypotheses evaluation in the context of a one way analysis of variance is illustrated using one of the studies from the OSF reproducibility project psychology \citep{OSF:2015}. \cite{Monin:2008} investigate the attraction to ``moral rebels'', that is, persons that take an unpopular but morally laudable stand. There are three groups in their experiment: in Group 1 participants rate their attraction to ``a person that is obedient and selects an African American person from a police line up of three''; in Group 2 participants execute a self-affirmation task intended to boost their self-confidence after which they rate ``a moral rebel who does not select the African American person''; and, in Group 3 participants execute a bogus writing task after which they rate ``a moral rebel''. The authors expect that the attraction to moral rebels is higher in the group executing the self-affirmation task (that boosts the confidence of the participants in that group) than in the group executing the bogus writing task, possibly even higher than in the group that rates the attraction of the obedient person. Their data will henceforth be referred to as the Monin data. Corresponding to their study are the following competing constrained models:
\begin{description}
\item $M_1: \mu_1 = \mu_2 = \mu_3$
\item $M_2: \mu_2 > \mu_1 > \mu_3$
\item $M_3: \text{neither $M_1$, nor $M_2$}$,
\end{description}
where, $\mu_1$, $\mu_2$, and $\mu_3$ denote the mean attractiveness scores in Groups 1, 2, and 3, respectively. Note that model $M_3$ denotes the complement model encompasses the subspace of $\mathbb{R}^3$ for $\bm\mu$ that does satisfy the constraints of $M_1$ and $M_2$.

In Table \ref{monin1} and \ref{monin2} the main results are presented. For model with no equality constraints or order constraints, the measures of relative fit and complexity omitted. The posterior model probabilities were computed using equal prior model probabilities (i.e., $P(M_t|\textbf{Y})=B_{tu}/\sum_{t'}B_{t'u}$). Note that the sufficient statistics correspond to the data reported by Monin, Sawyer, and Marques (2006).
As can be seen, for the three models under consideration, the order-constrained model $M_2$ receives is the best with a posterior probability of .963, followed by the complement model $M_3$, with a posterior probability of .036, and finally the equality-constrained null model received least evidence with a posterior probability of .001. This can be interpreted as very strong evidence for the order-constrained model. Note that the estimates and standard errors presented in Table \ref{monin1} also indicate evidence for the order-constrained model. Further note that the default prior probability that the order constraints of $M_2$ and $M_3$ hold under the unconstrained model equal $c_2^O=\frac{1}{6}$ and $c_3^O=\frac{5}{6}$, which are exactly equal to the probabilities that the order constraints hold under the unconstrained model when assuming that each ordering is equally likely a priori. This is a direct consequence of centering the unconstrained prior on the boundary of the order-constrained space. The presented default Bayes factors and posterior model probabilities confirm this suspicion by providing strong evidence in favor of the order-constrained model against the competing models.

\begin{table}[t]
\centering \caption{Unconstrained estimates for the ``Monin'' application.}\label{monin1}
\begin{tabular}{ccccccccc}
\hline
parameter & estimate & standard error & N \\
\hline
$\mu_1$ &1.88& .464& 19 \\
$\mu_2$ & 2.54& .464& 19\\
$\mu_3$ & 0.02& .375& 29\\
\hline
\end{tabular}
\end{table}

\begin{table}[t]
\centering \caption{Bayesian model selection for the ``Monin'' application.}\label{monin2}
\begin{tabular}{ccccccccc}
\hline
Model & $f_t^E$ & $c_t^E$ & $f_t^O$ & $c_t^O$ & $B_{tu}$ & $P(M_t|\textbf{y})$ \\
\hline
$M_1$ & 5.42e$-5$ & 8.45e$-3$ &  &  &  .006 & .001 \\
$M_2$ &  &  & .842 & .167 &  5.05 & .963 \\
$M_3$ &  &  & .158 & .833 &  .189 & .036 \\
\hline
\end{tabular}
\end{table}

\subsection{Multivariate multiple regression}
\cite{Stevens:1996} (Appendix A) presented data concerning the effect of the first year of the Sesame street series on the knowledge of 240 children in the age range 34 to 69 months. To illustrate informative hypothesis evaluation in the context of a multivariate multiple regression, the outcome variables $y_1$ and $y_2$, which are the knowledge of numbers and the knowledge of letters of children after watching Sesame Street, respectively, are regressed on $x_1$ and $x_2$, which are the knowledge of numbers and the knowledge of letters of children before watching Sesame Street for a year. In this application all data are standardized. The following multivariate multivariate multiple regression model will be used for $i=1,..,N,$ where $N=240$ denotes the sample size:
\begin{displaymath}
y_{i1} = \mu_{11} + \beta_{11} x_{i1} + \beta_{21} x_{i2} + e_{i1}
\end{displaymath}
\begin{equation}
y_{i2} = \mu_{12} + \beta_{12} x_{i2} + \beta_{22} x_{i2} + e_{i2}
\end{equation}
\begin{displaymath}
\left[\begin{array}{c}
e_{i1}\\
e_{i2}\\
\end{array}\right] \sim \mathcal{N}\left(\left[\begin{array}{c}
0\\
0\\
\end{array}\right],
\left[\begin{array}{cc}
\sigma^2_{1}& \sigma_{12}\\
\sigma_{12}& \sigma^2_{2}\\
\end{array}\right]\right).
\end{displaymath}
In this context expectations were formulated on the effects \textit{within} each dimension (letter knowledge and number knowledge), and on the effects \textit{between} the two dimensions. Within the knowledge dimensions, two competing expectations were formulated. First, it was expected that letter knowledge after watching Sesame Street can better be predicted by letter knowledge before watching Sesame Street than by number knowledge before watching Sesame Street. A similar expectation can be formulated for the number knowledge dimension. Furthermore it was expected that all effects were positive, i.e., $\beta_{11} > \beta_{21} > 0~ \&~ \beta_{22} > \beta_{12} > 0$. Second, it was expected that there was no effect of number knowledge before watching Sesame Street on letter knowledge after watching Sesame Street, and no effect of letter knowledge before watching Sesame Street on number knowledge after watching Sesame Street. The other effects were assumed positive, i.e.,
$\beta_{11} > \beta_{21} = 0~ \&~ \beta_{22} > \beta_{12} = 0$. Between the knowledge dimensions, it was expected that the effect of number knowledge of the pre-measurement on the post-measurement was equal, smaller, or larger than the effect of letter knowledge of the pre-measurement on the post-measurement, i.e.,
$\beta_{11}=\beta_{22}$ or $\beta_{11}<\beta_{22}$ or $\beta_{11}>\beta_{22}$.

Combining these different expectations we can formulate 7 competing constrained models:
\begin{align*}
M_1&:\beta_{11} > \beta_{21} > 0~ \&~ \beta_{22} > \beta_{12} > 0~ \&~ \beta_{11}=\beta_{22}\\
M_2&:\beta_{11} > \beta_{21} > 0~ \&~ \beta_{22} > \beta_{12} > 0~ \&~ \beta_{11}<\beta_{22}\\
M_3&:\beta_{11} > \beta_{21} > 0~ \&~ \beta_{22} > \beta_{12} > 0~ \&~ \beta_{11}>\beta_{22}\\
M_4&:\beta_{11} > \beta_{21} = 0~ \&~ \beta_{22} > \beta_{12} = 0~ \&~ \beta_{11}=\beta_{22}\\
M_5&:\beta_{11} > \beta_{21} = 0~ \&~ \beta_{22} > \beta_{12} = 0~ \&~ \beta_{11}<\beta_{22}\\
M_6&:\beta_{11} > \beta_{21} = 0~ \&~ \beta_{22} > \beta_{12} = 0~ \&~ \beta_{11}>\beta_{22}\\
M_7&:\text{not $M_1,\ldots,$ or $M_6$}.
\end{align*}

\begin{table}[t]
\centering \caption{Unconstrained estimates for the Sesame Street application.}\label{sesam1}
\begin{tabular}{ccclccccc}
\hline
parameter & estimate & standard error & ~~~~~~~~correlation matrix\\
\hline
$\beta_{11}$  & .647 & .069 & ~~1.00\\
$\beta_{21}$  & .040 & .069 & ~-.717 ~~~ 1.00 \\
$\beta_{12}$  & .428 & .073 & ~~.708  ~~ -.508 ~~ 1.00 \\
$\beta_{22}$  & .242 & .073 & ~-.508 ~~~ .708 ~ -.717 ~~ 1.00\\
\hline
\end{tabular}
\end{table}

\begin{table}[t]
\centering \caption{Bayesian Hypothesis Evaluation for the Sesame Street application.}\label{sesam2}
\begin{tabular}{ccccccccc}
\hline\hline
Model & $f_t^E$ & $c_t^E$ & $f_t^O$ & $c_t^O$ &   $B_{tu}$ & $P(M_t|\textbf{y})$  \\
\hline
$M_1$ & .016 & .161 & 1.00 & .037 & 2.71  & .111   \\
$M_2$ &      &      & .000 & .006 & .063  & .003   \\
$M_3$ &      &      & .082 & .004 & 20.8  & .849   \\
$M_4$ & .000 & .112 & 1.00 & .500 & .000  & .000   \\
$M_5$ & .000 & .130 & .111 & .171 & .000  & .000   \\
$M_6$ & .000 & .130 & .888 & .156 & .000  & .000   \\
$M_7$ &      &      & .918 & .991 & .926  & .038   \\
\hline
\end{tabular}
\end{table}

The MLEs with standard errors and correlations, and the Bayes factors and posterior model probabilities (assuming equal prior model probabilities) together with the measures of relative fit and complexity (if available) are presented in Table \ref{sesam1} and \ref{sesam2}, respectively. The Bayes factors and posterior model probabilities show that the data provides most evidence for the order-constrained model $M_3$ with a posterior probability of .849, while the constrained model $M_1$ and the complement model $M_7$ also receive some mild evidence with posterior probabilities of .111 and .038, respectively. The other models can essentially be ruled out given the very low posterior model probabilities when assuming that all models were equally likely before observing the data (as we assumed in the computation). Interestingly the unconstrained estimates of $(\beta_{22},\beta_{11})$, namely $(.647,.242)$, are not in agreement with the constraint $\beta_{22}>\beta_{11}$ under model $M_3$. The reason that there is still most evidence for $M_3$ is because the posterior probability that the constraints of $M_3$ hold under $M_u$, i.e., $f^O_3\approx.082$, is about 20.8 times larger than the default prior probability that the constraints hold, i.e., $c^O_3\approx .004$. As the Bayes factor quantifies the change in support prior to posterior \citep{Lavine:1999}, these prior and posterior probabilities suggest there is evidence for $M_3$ after observing the data. Further note that the Bayes factor functions as an Occam's razor by balancing between model complexity and model fit \citep{Berger:1999,Mulder:2010}. As the posterior probability of .082 can be seen as a measure of relative fit for $M_3$ and the prior probability of .004 as a measure of relative complexity for $M_3$ relative to $H_u$, this implies that $M_3$ has the best ratio of fit and complexity out of all models under evaluation, resulting in most evidence for $M_3$.



\section{Concluding remarks}
A default Bayes factor was proposed for evaluating multivariate normal linear models with competing sets of equality and order constraints on the parameters of interest. The methodology has the following attractive features. First the method can be used for evaluating statistics models with equality as well as order constraints on the parameters of interest. The possibility of order constrained testing is particularly useful in the applied sciences where researchers often formulate their scientific expectations using order constraints. Second, the method is fully automatic and therefore can be applied when prior information is weak or completely unavailable. The default prior is based on a minimal fraction of the information in the observed data of every group so that maximal information is used for model selection. Third, the Bayes factor is relatively simple to compute via Monte Carlo estimation that can be done in parallel. The Bayes factor has analytic expressions for special cases. Fourth the criterion is consistent which implies that the true constrained model will always be selected it the sample is large enough. Fifth, in the presence of missing data that are missing at random, the Bayes factor can be computed relatively easily using a multiple imputation method only under the unconstrained model. In sum, the method gives substantive researchers a simple tool for quantifying the evidence between competing scientific expectations, updating the evidence as new data emerge, while also correcting for missing data that are missing at random for many popular models including (multivariate) linear regression, (M)AN(C)OVA, repeated measures. The methodology will be implemented in the R-package `BFpack' that is scheduled for later this year.

In this paper the Bayes factor was used as a confirmatory tool for model selection among a specific set of models with equality and/or order constraints. Equal model prior model probabilities were considered because all models were (approximately) equally plausible based on substantive justifications. In a more exploratory setting other choices may be preferable, \cite[e.g., see][who considered a model selection problem of many competing equality constrained models]{Scott:2006}. It will be interesting to investigate how prior model probabilities should be specified in such exploratory settings when models may contain equality as well as order constraints on the parameters of interest.

\section*{Acknowledgements}
The first author is supported by a Vidi Grant (452-17-006) awarded by the Netherlands Organization for Scientific Research (NWO). The second author is supported by the Consortium on Individual Development (CID) which is funded through the Gravitation program of the Dutch Ministry of Education, Culture, and Science and NWO (024.001.003).

\bibliographystyle{apacite}
\bibliography{refs_mulder}

\begin{thebibliography}{}

\bibitem [\protect \citeauthoryear {%
Bandekar%
\ \BBA {} Nagar%
}{%
Bandekar%
\ \BBA {} Nagar%
}{%
{\protect \APACyear {2003}}%
}]{%
BandekarNagar:2003}
\APACinsertmetastar {%
BandekarNagar:2003}%
\begin{APACrefauthors}%
Bandekar, R\BPBI R.%
\BCBT {}\ \BBA {} Nagar, D\BPBI K.%
\end{APACrefauthors}%
\unskip\
\newblock
\APACrefYearMonthDay{2003}{}{}.
\newblock
{\BBOQ}\APACrefatitle {Matrix variate {C}auchy distribution} {Matrix variate
  {C}auchy distribution}.{\BBCQ}
\newblock
\APACjournalVolNumPages{Statistics: A Journal of Theoretical and Applied
  Statistics}{37}{}{537--550}.
\PrintBackRefs{\CurrentBib}

\bibitem [\protect \citeauthoryear {%
Berger%
}{%
Berger%
}{%
{\protect \APACyear {2006}}%
}]{%
Berger:2006}
\APACinsertmetastar {%
Berger:2006}%
\begin{APACrefauthors}%
Berger, J\BPBI O.%
\end{APACrefauthors}%
\unskip\
\newblock
\APACrefYearMonthDay{2006}{}{}.
\newblock
{\BBOQ}\APACrefatitle {The case for objective {B}ayesian analysis} {The case
  for objective {B}ayesian analysis}.{\BBCQ}
\newblock
\APACjournalVolNumPages{Bayesian Analysis}{1}{}{385--402}.
\PrintBackRefs{\CurrentBib}

\bibitem [\protect \citeauthoryear {%
Berger%
\ \BBA {} Mortera%
}{%
Berger%
\ \BBA {} Mortera%
}{%
{\protect \APACyear {1995}}%
}]{%
Berger:1995}
\APACinsertmetastar {%
Berger:1995}%
\begin{APACrefauthors}%
Berger, J\BPBI O.%
\BCBT {}\ \BBA {} Mortera, J.%
\end{APACrefauthors}%
\unskip\
\newblock
\APACrefYearMonthDay{1995}{}{}.
\newblock
{\BBOQ}\APACrefatitle {Discussion to fractional Bayes factors for model
  comparison (by {O}'{H}agan)} {Discussion to fractional bayes factors for
  model comparison (by {O}'{H}agan)}.{\BBCQ}
\newblock
\APACjournalVolNumPages{Journal of the Royal Statistical Society Series
  B}{56}{}{130}.
\PrintBackRefs{\CurrentBib}

\bibitem [\protect \citeauthoryear {%
Berger%
\ \BBA {} Mortera%
}{%
Berger%
\ \BBA {} Mortera%
}{%
{\protect \APACyear {1999}}%
}]{%
Berger:1999}
\APACinsertmetastar {%
Berger:1999}%
\begin{APACrefauthors}%
Berger, J\BPBI O.%
\BCBT {}\ \BBA {} Mortera, J.%
\end{APACrefauthors}%
\unskip\
\newblock
\APACrefYearMonthDay{1999}{}{}.
\newblock
{\BBOQ}\APACrefatitle {Default {B}ayes factors for nonnested hypothesis
  testing} {Default {B}ayes factors for nonnested hypothesis testing}.{\BBCQ}
\newblock
\APACjournalVolNumPages{Journal of American Statistical
  Association}{94}{}{542--554}.
\PrintBackRefs{\CurrentBib}

\bibitem [\protect \citeauthoryear {%
Berger%
\ \BBA {} Pericchi%
}{%
Berger%
\ \BBA {} Pericchi%
}{%
{\protect \APACyear {1996}}%
}]{%
Berger:1996}
\APACinsertmetastar {%
Berger:1996}%
\begin{APACrefauthors}%
Berger, J\BPBI O.%
\BCBT {}\ \BBA {} Pericchi, L\BPBI R.%
\end{APACrefauthors}%
\unskip\
\newblock
\APACrefYearMonthDay{1996}{}{}.
\newblock
{\BBOQ}\APACrefatitle {The intrinsic {B}ayes factor for model selection and
  prediction} {The intrinsic {B}ayes factor for model selection and
  prediction}.{\BBCQ}
\newblock
\APACjournalVolNumPages{Journal of the American Statistical
  Association}{91}{}{109--122}.
\PrintBackRefs{\CurrentBib}

\bibitem [\protect \citeauthoryear {%
Berger%
\ \BBA {} Pericchi%
}{%
Berger%
\ \BBA {} Pericchi%
}{%
{\protect \APACyear {2004}}%
}]{%
Berger:2004}
\APACinsertmetastar {%
Berger:2004}%
\begin{APACrefauthors}%
Berger, J\BPBI O.%
\BCBT {}\ \BBA {} Pericchi, L\BPBI R.%
\end{APACrefauthors}%
\unskip\
\newblock
\APACrefYearMonthDay{2004}{}{}.
\newblock
{\BBOQ}\APACrefatitle {Training Samples in Objective {B}ayesian Model
  Selection} {Training samples in objective {B}ayesian model selection}.{\BBCQ}
\newblock
\APACjournalVolNumPages{The Annals of Statistics}{32}{3}{841--869}.
\PrintBackRefs{\CurrentBib}

\bibitem [\protect \citeauthoryear {%
{B\"{o}ing-Messing}%
, van Assen%
, Hofman%
, Hoijtink%
\BCBL {}\ \BBA {} Mulder%
}{%
{B\"{o}ing-Messing}%
\ \protect \BOthers {.}}{%
{\protect \APACyear {2017}}%
}]{%
BoingMessing:2017}
\APACinsertmetastar {%
BoingMessing:2017}%
\begin{APACrefauthors}%
{B\"{o}ing-Messing}, F.%
, van Assen, M.%
, Hofman, A.%
, Hoijtink, H.%
\BCBL {}\ \BBA {} Mulder, J.%
\end{APACrefauthors}%
\unskip\
\newblock
\APACrefYearMonthDay{2017}{}{}.
\newblock
{\BBOQ}\APACrefatitle {Bayesian Evaluation of Constrained Hypotheses on
  Variances of Multiple Independent Groups} {Bayesian evaluation of constrained
  hypotheses on variances of multiple independent groups}.{\BBCQ}
\newblock
\APACjournalVolNumPages{Psychological Methods}{22}{}{262--287}.
\PrintBackRefs{\CurrentBib}

\bibitem [\protect \citeauthoryear {%
Box%
\ \BBA {} Tiao%
}{%
Box%
\ \BBA {} Tiao%
}{%
{\protect \APACyear {1973}}%
}]{%
BoxTiao}
\APACinsertmetastar {%
BoxTiao}%
\begin{APACrefauthors}%
Box, G\BPBI E\BPBI P.%
\BCBT {}\ \BBA {} Tiao, G\BPBI C.%
\end{APACrefauthors}%
\unskip\
\newblock
\APACrefYear{1973}.
\newblock
\APACrefbtitle {Bayesian Inference in Statistical Snalysis} {Bayesian inference
  in statistical snalysis}.
\newblock
\APACaddressPublisher{}{Reading, MA: Addison-Wesley}.
\PrintBackRefs{\CurrentBib}

\bibitem [\protect \citeauthoryear {%
Braeken%
, Mulder%
\BCBL {}\ \BBA {} Wood%
}{%
Braeken%
\ \protect \BOthers {.}}{%
{\protect \APACyear {2015}}%
}]{%
Braeken:2015}
\APACinsertmetastar {%
Braeken:2015}%
\begin{APACrefauthors}%
Braeken, J.%
, Mulder, J.%
\BCBL {}\ \BBA {} Wood, S.%
\end{APACrefauthors}%
\unskip\
\newblock
\APACrefYearMonthDay{2015}{}{}.
\newblock
{\BBOQ}\APACrefatitle {Relative effects at work: {B}ayes factors for order
  hypotheses} {Relative effects at work: {B}ayes factors for order
  hypotheses}.{\BBCQ}
\newblock
\APACjournalVolNumPages{Journal of Management}{41}{}{}.
\PrintBackRefs{\CurrentBib}

\bibitem [\protect \citeauthoryear {%
Conigliani%
\ \BBA {} O'Hagan%
}{%
Conigliani%
\ \BBA {} O'Hagan%
}{%
{\protect \APACyear {2000}}%
}]{%
Conigliani:2000}
\APACinsertmetastar {%
Conigliani:2000}%
\begin{APACrefauthors}%
Conigliani, C.%
\BCBT {}\ \BBA {} O'Hagan, A.%
\end{APACrefauthors}%
\unskip\
\newblock
\APACrefYearMonthDay{2000}{}{}.
\newblock
{\BBOQ}\APACrefatitle {Sensitivity of the fractional {B}ayes factor to prior
  distributions} {Sensitivity of the fractional {B}ayes factor to prior
  distributions}.{\BBCQ}
\newblock
\APACjournalVolNumPages{The Canadian Journal of Statistics}{28}{}{343--352}.
\PrintBackRefs{\CurrentBib}

\bibitem [\protect \citeauthoryear {%
Consonni%
\ \BBA {} Paroli%
}{%
Consonni%
\ \BBA {} Paroli%
}{%
{\protect \APACyear {2017}}%
}]{%
Consonni:2017}
\APACinsertmetastar {%
Consonni:2017}%
\begin{APACrefauthors}%
Consonni, G.%
\BCBT {}\ \BBA {} Paroli, R.%
\end{APACrefauthors}%
\unskip\
\newblock
\APACrefYearMonthDay{2017}{}{}.
\newblock
{\BBOQ}\APACrefatitle {Objective {B}ayesian comparison of constrained analysis
  of variance models} {Objective {B}ayesian comparison of constrained analysis
  of variance models}.{\BBCQ}
\newblock
\APACjournalVolNumPages{Psychometrika}{}{}{}.
\PrintBackRefs{\CurrentBib}

\bibitem [\protect \citeauthoryear {%
{De Santis}%
\ \BBA {} Spezzaferri%
}{%
{De Santis}%
\ \BBA {} Spezzaferri%
}{%
{\protect \APACyear {1997}}%
}]{%
DeSantis:1997}
\APACinsertmetastar {%
DeSantis:1997}%
\begin{APACrefauthors}%
{De Santis}, F.%
\BCBT {}\ \BBA {} Spezzaferri, F.%
\end{APACrefauthors}%
\unskip\
\newblock
\APACrefYearMonthDay{1997}{}{}.
\newblock
{\BBOQ}\APACrefatitle {Alternative {B}ayes factors for model selection}
  {Alternative {B}ayes factors for model selection}.{\BBCQ}
\newblock
\APACjournalVolNumPages{Canadian Journal of Statistics}{25}{}{503--515}.
\PrintBackRefs{\CurrentBib}

\bibitem [\protect \citeauthoryear {%
{De Santis}%
\ \BBA {} Spezzaferri%
}{%
{De Santis}%
\ \BBA {} Spezzaferri%
}{%
{\protect \APACyear {2001}}%
}]{%
DeSantis:2001}
\APACinsertmetastar {%
DeSantis:2001}%
\begin{APACrefauthors}%
{De Santis}, F.%
\BCBT {}\ \BBA {} Spezzaferri, F.%
\end{APACrefauthors}%
\unskip\
\newblock
\APACrefYearMonthDay{2001}{}{}.
\newblock
{\BBOQ}\APACrefatitle {Consistent fractional Bayes factor for nested normal
  linear models} {Consistent fractional bayes factor for nested normal linear
  models}.{\BBCQ}
\newblock
\APACjournalVolNumPages{Journal of Statistical Planning and
  Inference}{97}{}{305--321}.
\PrintBackRefs{\CurrentBib}

\bibitem [\protect \citeauthoryear {%
de Jong%
, Rigotti%
\BCBL {}\ \BBA {} Mulder%
}{%
de Jong%
\ \protect \BOthers {.}}{%
{\protect \APACyear {2017}}%
}]{%
DeJong:2017}
\APACinsertmetastar {%
DeJong:2017}%
\begin{APACrefauthors}%
de Jong, J.%
, Rigotti, T.%
\BCBL {}\ \BBA {} Mulder, J.%
\end{APACrefauthors}%
\unskip\
\newblock
\APACrefYearMonthDay{2017}{}{}.
\newblock
{\BBOQ}\APACrefatitle {One after the other: Effects of sequence patterns of
  breached and overfulfilled obligations} {One after the other: Effects of
  sequence patterns of breached and overfulfilled obligations}.{\BBCQ}
\newblock
\APACjournalVolNumPages{European Journal of Work and Organizational
  Psychology}{26}{}{337--355}.
\PrintBackRefs{\CurrentBib}

\bibitem [\protect \citeauthoryear {%
Dickey%
}{%
Dickey%
}{%
{\protect \APACyear {1971}}%
}]{%
Dickey:1971}
\APACinsertmetastar {%
Dickey:1971}%
\begin{APACrefauthors}%
Dickey, J.%
\end{APACrefauthors}%
\unskip\
\newblock
\APACrefYearMonthDay{1971}{}{}.
\newblock
{\BBOQ}\APACrefatitle {The weighted likelihood ratio, linear hypotheses on
  normal location parameters} {The weighted likelihood ratio, linear hypotheses
  on normal location parameters}.{\BBCQ}
\newblock
\APACjournalVolNumPages{The Annals of Statistics}{42}{}{204--223}.
\PrintBackRefs{\CurrentBib}

\bibitem [\protect \citeauthoryear {%
Dogge%
, Gayet%
, Custers%
, Hoijtink%
\BCBL {}\ \BBA {} Aarts%
}{%
Dogge%
\ \protect \BOthers {.}}{%
{\protect \APACyear {2019}}%
}]{%
Digge:2019}
\APACinsertmetastar {%
Digge:2019}%
\begin{APACrefauthors}%
Dogge, M.%
, Gayet, S.%
, Custers, R.%
, Hoijtink, H.%
\BCBL {}\ \BBA {} Aarts, H.%
\end{APACrefauthors}%
\unskip\
\newblock
\APACrefYearMonthDay{2019}{}{}.
\newblock
{\BBOQ}\APACrefatitle {Perception of action-outcomes is shaped by life-long and
  contextual expectations} {Perception of action-outcomes is shaped by
  life-long and contextual expectations}.{\BBCQ}
\newblock
\APACjournalVolNumPages{Scientific Reports}{}{}{}.
\PrintBackRefs{\CurrentBib}

\bibitem [\protect \citeauthoryear {%
Flore%
, Mulder%
\BCBL {}\ \BBA {} Wicherts%
}{%
Flore%
\ \protect \BOthers {.}}{%
{\protect \APACyear {2019}}%
}]{%
Flore:2019}
\APACinsertmetastar {%
Flore:2019}%
\begin{APACrefauthors}%
Flore, P\BPBI C.%
, Mulder, J.%
\BCBL {}\ \BBA {} Wicherts, J\BPBI M.%
\end{APACrefauthors}%
\unskip\
\newblock
\APACrefYearMonthDay{2019}{}{}.
\newblock
{\BBOQ}\APACrefatitle {The influence of gender stereotype threat on mathematics
  test scores of Dutch high school students: a registered report} {The
  influence of gender stereotype threat on mathematics test scores of dutch
  high school students: a registered report}.{\BBCQ}
\newblock
\APACjournalVolNumPages{Comprehensive Results in Social Psychology}{}{}{}.
\PrintBackRefs{\CurrentBib}

\bibitem [\protect \citeauthoryear {%
Genz%
\ \protect \BOthers {.}}{%
Genz%
\ \protect \BOthers {.}}{%
{\protect \APACyear {2016}}%
}]{%
Genz:2016}
\APACinsertmetastar {%
Genz:2016}%
\begin{APACrefauthors}%
Genz, A.%
, Bretz, F.%
, Miwa, T.%
, Mi, X.%
, Leisch, F.%
, Scheipl, F.%
\BDBL {}Hothorn, T.%
\end{APACrefauthors}%
\unskip\
\newblock
\APACrefYearMonthDay{2016}{}{}.
\newblock
{\BBOQ}\APACrefatitle {R-package `mvtnorm'} {R-package `mvtnorm'}{\BBCQ}\
  [\bibcomputersoftwaremanual].
\newblock
\APACrefnote{R package version 1.14.4 --- For new features, see the 'Changelog'
  file (in the package source)}
\PrintBackRefs{\CurrentBib}

\bibitem [\protect \citeauthoryear {%
Gilks%
}{%
Gilks%
}{%
{\protect \APACyear {1995}}%
}]{%
Gilks:1995}
\APACinsertmetastar {%
Gilks:1995}%
\begin{APACrefauthors}%
Gilks, W\BPBI R.%
\end{APACrefauthors}%
\unskip\
\newblock
\APACrefYearMonthDay{1995}{}{}.
\newblock
{\BBOQ}\APACrefatitle {Discussion to fractional Bayes factors for model
  comparison (by {O}'{H}agan)} {Discussion to fractional bayes factors for
  model comparison (by {O}'{H}agan)}.{\BBCQ}
\newblock
\APACjournalVolNumPages{Journal of the Royal Statistical Society Series
  B}{56}{}{118--120}.
\PrintBackRefs{\CurrentBib}

\bibitem [\protect \citeauthoryear {%
Gu%
, Hoijtink%
, Mulder%
\BCBL {}\ \BBA {} Rosseel%
}{%
Gu%
\ \protect \BOthers {.}}{%
{\protect \APACyear {2019}}%
}]{%
Gu:2019}
\APACinsertmetastar {%
Gu:2019}%
\begin{APACrefauthors}%
Gu, X.%
, Hoijtink, H.%
, Mulder, J.%
\BCBL {}\ \BBA {} Rosseel, Y.%
\end{APACrefauthors}%
\unskip\
\newblock
\APACrefYearMonthDay{2019}{}{}.
\newblock
{\BBOQ}\APACrefatitle {Bain: A program for the evaluation of inequality
  constrained hypotheses using {B}ayes factors in structural equation models}
  {Bain: A program for the evaluation of inequality constrained hypotheses
  using {B}ayes factors in structural equation models}.{\BBCQ}
\newblock
\APACjournalVolNumPages{Journal of Statistical Computation and
  Simulation}{}{}{}.
\PrintBackRefs{\CurrentBib}

\bibitem [\protect \citeauthoryear {%
Gu%
, Mulder%
, Decovic%
\BCBL {}\ \BBA {} Hoijtink%
}{%
Gu%
\ \protect \BOthers {.}}{%
{\protect \APACyear {2014}}%
}]{%
Gu:2014}
\APACinsertmetastar {%
Gu:2014}%
\begin{APACrefauthors}%
Gu, X.%
, Mulder, J.%
, Decovic, M.%
\BCBL {}\ \BBA {} Hoijtink, H.%
\end{APACrefauthors}%
\unskip\
\newblock
\APACrefYearMonthDay{2014}{}{}.
\newblock
{\BBOQ}\APACrefatitle {Bayesian evaluation of inequality constrained
  hypotheses} {Bayesian evaluation of inequality constrained
  hypotheses}.{\BBCQ}
\newblock
\APACjournalVolNumPages{Psychological Methods}{19}{}{511--527}.
\PrintBackRefs{\CurrentBib}

\bibitem [\protect \citeauthoryear {%
Gu%
, Mulder%
\BCBL {}\ \BBA {} Hoijtink%
}{%
Gu%
\ \protect \BOthers {.}}{%
{\protect \APACyear {2017}}%
}]{%
Gu:2017}
\APACinsertmetastar {%
Gu:2017}%
\begin{APACrefauthors}%
Gu, X.%
, Mulder, J.%
\BCBL {}\ \BBA {} Hoijtink, H.%
\end{APACrefauthors}%
\unskip\
\newblock
\APACrefYearMonthDay{2017}{}{}.
\newblock
{\BBOQ}\APACrefatitle {Approximated adjusted fractional Bayes factors: A
  general method for testing informative hypotheses} {Approximated adjusted
  fractional bayes factors: A general method for testing informative
  hypotheses}.{\BBCQ}
\newblock
\APACjournalVolNumPages{British Journal of Mathematical and Statistical
  Psychology}{}{}{}.
\PrintBackRefs{\CurrentBib}

\bibitem [\protect \citeauthoryear {%
Hoijtink%
}{%
Hoijtink%
}{%
{\protect \APACyear {2011}}%
}]{%
Hoijtink:2011}
\APACinsertmetastar {%
Hoijtink:2011}%
\begin{APACrefauthors}%
Hoijtink, H.%
\end{APACrefauthors}%
\unskip\
\newblock
\APACrefYear{2011}.
\newblock
\APACrefbtitle {Informative Hypotheses: Theory and Practice for Behavioral and
  Social Scientists} {Informative hypotheses: Theory and practice for
  behavioral and social scientists}.
\newblock
\APACaddressPublisher{}{New York: Chapman \& Hall/CRC}.
\PrintBackRefs{\CurrentBib}

\bibitem [\protect \citeauthoryear {%
Hoijtink%
, Gu%
\BCBL {}\ \BBA {} Mulder%
}{%
Hoijtink%
, Gu%
\BCBL {}\ \BBA {} Mulder%
}{%
{\protect \APACyear {2018}}%
}]{%
Hoijtink:2018}
\APACinsertmetastar {%
Hoijtink:2018}%
\begin{APACrefauthors}%
Hoijtink, H.%
, Gu, X.%
\BCBL {}\ \BBA {} Mulder, J.%
\end{APACrefauthors}%
\unskip\
\newblock
\APACrefYearMonthDay{2018}{}{}.
\newblock
{\BBOQ}\APACrefatitle {Bayesian evaluation of informative hypotheses for
  multiple populations} {Bayesian evaluation of informative hypotheses for
  multiple populations}.{\BBCQ}
\newblock
\APACjournalVolNumPages{British Journal of Mathematical and Statistical
  Psychology}{}{}{}.
\PrintBackRefs{\CurrentBib}

\bibitem [\protect \citeauthoryear {%
Hoijtink%
, Gu%
, Mulder%
\BCBL {}\ \BBA {} Rosseel%
}{%
Hoijtink%
, Gu%
, Mulder%
\BCBL {}\ \BBA {} Rosseel%
}{%
{\protect \APACyear {2018}}%
}]{%
Hoijtink:2018b}
\APACinsertmetastar {%
Hoijtink:2018b}%
\begin{APACrefauthors}%
Hoijtink, H.%
, Gu, X.%
, Mulder, J.%
\BCBL {}\ \BBA {} Rosseel, Y.%
\end{APACrefauthors}%
\unskip\
\newblock
\APACrefYearMonthDay{2018}{}{}.
\newblock
{\BBOQ}\APACrefatitle {Computing {B}ayes Factors From Data With Missing Values}
  {Computing {B}ayes factors from data with missing values}.{\BBCQ}
\newblock
\APACjournalVolNumPages{Psychological Methods}{}{}{}.
\PrintBackRefs{\CurrentBib}

\bibitem [\protect \citeauthoryear {%
Jeffreys%
}{%
Jeffreys%
}{%
{\protect \APACyear {1961}}%
}]{%
Jeffreys}
\APACinsertmetastar {%
Jeffreys}%
\begin{APACrefauthors}%
Jeffreys, H.%
\end{APACrefauthors}%
\unskip\
\newblock
\APACrefYear{1961}.
\newblock
\APACrefbtitle {Theory of Probability-3rd ed} {Theory of probability-3rd ed}.
\newblock
\APACaddressPublisher{}{New York: Oxford University Press}.
\PrintBackRefs{\CurrentBib}

\bibitem [\protect \citeauthoryear {%
Kass%
\ \BBA {} Raftery%
}{%
Kass%
\ \BBA {} Raftery%
}{%
{\protect \APACyear {1995}}%
}]{%
Kass:1995}
\APACinsertmetastar {%
Kass:1995}%
\begin{APACrefauthors}%
Kass, R\BPBI E.%
\BCBT {}\ \BBA {} Raftery, A\BPBI E.%
\end{APACrefauthors}%
\unskip\
\newblock
\APACrefYearMonthDay{1995}{}{}.
\newblock
{\BBOQ}\APACrefatitle {{B}ayes Factors} {{B}ayes factors}.{\BBCQ}
\newblock
\APACjournalVolNumPages{Journal of American Statistical
  Association}{90}{}{773--795}.
\PrintBackRefs{\CurrentBib}

\bibitem [\protect \citeauthoryear {%
Klugkist%
, Laudy%
\BCBL {}\ \BBA {} Hoijtink%
}{%
Klugkist%
\ \protect \BOthers {.}}{%
{\protect \APACyear {2005}}%
}]{%
Klugkist:2005}
\APACinsertmetastar {%
Klugkist:2005}%
\begin{APACrefauthors}%
Klugkist, I.%
, Laudy, O.%
\BCBL {}\ \BBA {} Hoijtink, H.%
\end{APACrefauthors}%
\unskip\
\newblock
\APACrefYearMonthDay{2005}{}{}.
\newblock
{\BBOQ}\APACrefatitle {Inequality constrained analysis of variance: A
  {B}ayesian approach} {Inequality constrained analysis of variance: A
  {B}ayesian approach}.{\BBCQ}
\newblock
\APACjournalVolNumPages{Psychological Methods}{10}{}{477--493}.
\PrintBackRefs{\CurrentBib}

\bibitem [\protect \citeauthoryear {%
Klugkist%
, Laudy%
\BCBL {}\ \BBA {} Hoijtink%
}{%
Klugkist%
\ \protect \BOthers {.}}{%
{\protect \APACyear {2010}}%
}]{%
Klugkist:2010}
\APACinsertmetastar {%
Klugkist:2010}%
\begin{APACrefauthors}%
Klugkist, I.%
, Laudy, O.%
\BCBL {}\ \BBA {} Hoijtink, H.%
\end{APACrefauthors}%
\unskip\
\newblock
\APACrefYearMonthDay{2010}{}{}.
\newblock
{\BBOQ}\APACrefatitle {{B}ayesian evaluation of inequality and equality
  constrained hypotheses for contingency tables} {{B}ayesian evaluation of
  inequality and equality constrained hypotheses for contingency
  tables}.{\BBCQ}
\newblock
\APACjournalVolNumPages{Psychological Methods}{15}{}{281--299}.
\PrintBackRefs{\CurrentBib}

\bibitem [\protect \citeauthoryear {%
Kluytmans%
, {van de Schoot}%
, Mulder%
\BCBL {}\ \BBA {} Hoijtink%
}{%
Kluytmans%
\ \protect \BOthers {.}}{%
{\protect \APACyear {2012}}%
}]{%
Kluytmans:2012}
\APACinsertmetastar {%
Kluytmans:2012}%
\begin{APACrefauthors}%
Kluytmans, A.%
, {van de Schoot}, R.%
, Mulder, J.%
\BCBL {}\ \BBA {} Hoijtink, H.%
\end{APACrefauthors}%
\unskip\
\newblock
\APACrefYearMonthDay{2012}{}{}.
\newblock
{\BBOQ}\APACrefatitle {Illustrating Bayesian evaluation of informative
  hypotheses for regression models} {Illustrating bayesian evaluation of
  informative hypotheses for regression models}.{\BBCQ}
\newblock
\APACjournalVolNumPages{Frontiers in Psychology}{3}{}{1--11}.
\newblock
\begin{APACrefDOI} \doi{10.3389/fpsyg.2012.00002} \end{APACrefDOI}
\PrintBackRefs{\CurrentBib}

\bibitem [\protect \citeauthoryear {%
Lavine%
\ \BBA {} Chervish%
}{%
Lavine%
\ \BBA {} Chervish%
}{%
{\protect \APACyear {1999}}%
}]{%
Lavine:1999}
\APACinsertmetastar {%
Lavine:1999}%
\begin{APACrefauthors}%
Lavine, M.%
\BCBT {}\ \BBA {} Chervish, M\BPBI J.%
\end{APACrefauthors}%
\unskip\
\newblock
\APACrefYearMonthDay{1999}{}{}.
\newblock
{\BBOQ}\APACrefatitle {{B}ayes factors: {W}hat they are and what they are not}
  {{B}ayes factors: {W}hat they are and what they are not}.{\BBCQ}
\newblock
\APACjournalVolNumPages{The American Statistician}{53}{}{119--122}.
\PrintBackRefs{\CurrentBib}

\bibitem [\protect \citeauthoryear {%
Lindley%
}{%
Lindley%
}{%
{\protect \APACyear {1995}}%
}]{%
Lindley:1995}
\APACinsertmetastar {%
Lindley:1995}%
\begin{APACrefauthors}%
Lindley, D\BPBI V.%
\end{APACrefauthors}%
\unskip\
\newblock
\APACrefYearMonthDay{1995}{}{}.
\newblock
{\BBOQ}\APACrefatitle {Discussion to fractional Bayes factors for model
  comparison (by {O}'{H}agan)} {Discussion to fractional bayes factors for
  model comparison (by {O}'{H}agan)}.{\BBCQ}
\newblock
\APACjournalVolNumPages{Journal of the Royal Statistical Society Series
  B}{56}{}{123}.
\PrintBackRefs{\CurrentBib}

\bibitem [\protect \citeauthoryear {%
Monin%
, Sawyer%
\BCBL {}\ \BBA {} Marquez%
}{%
Monin%
\ \protect \BOthers {.}}{%
{\protect \APACyear {2008}}%
}]{%
Monin:2008}
\APACinsertmetastar {%
Monin:2008}%
\begin{APACrefauthors}%
Monin, B.%
, Sawyer, P.%
\BCBL {}\ \BBA {} Marquez, M.%
\end{APACrefauthors}%
\unskip\
\newblock
\APACrefYearMonthDay{2008}{}{}.
\newblock
{\BBOQ}\APACrefatitle {The rejection of moral rebels: {R}esenting those who do
  the right thing} {The rejection of moral rebels: {R}esenting those who do the
  right thing}.{\BBCQ}
\newblock
\APACjournalVolNumPages{Journal of Personality and Social
  Psychology}{95}{}{76--93}.
\PrintBackRefs{\CurrentBib}

\bibitem [\protect \citeauthoryear {%
Moreno%
, Bertolino%
\BCBL {}\ \BBA {} Racugno%
}{%
Moreno%
\ \protect \BOthers {.}}{%
{\protect \APACyear {1998}}%
}]{%
Moreno:1998}
\APACinsertmetastar {%
Moreno:1998}%
\begin{APACrefauthors}%
Moreno, E.%
, Bertolino, F.%
\BCBL {}\ \BBA {} Racugno, W.%
\end{APACrefauthors}%
\unskip\
\newblock
\APACrefYearMonthDay{1998}{}{}.
\newblock
{\BBOQ}\APACrefatitle {An intrinsic limiting procedure for model selection and
  hypotheses testing} {An intrinsic limiting procedure for model selection and
  hypotheses testing}.{\BBCQ}
\newblock
\APACjournalVolNumPages{Journal of the American Statistical
  Association}{93}{}{1451--1460}.
\PrintBackRefs{\CurrentBib}

\bibitem [\protect \citeauthoryear {%
Mulder%
}{%
Mulder%
}{%
{\protect \APACyear {2014}}%
}]{%
Mulder:2014b}
\APACinsertmetastar {%
Mulder:2014b}%
\begin{APACrefauthors}%
Mulder, J.%
\end{APACrefauthors}%
\unskip\
\newblock
\APACrefYearMonthDay{2014}{}{}.
\newblock
{\BBOQ}\APACrefatitle {Prior adjusted default {B}ayes factors for testing
  (in)equality constrained hypotheses} {Prior adjusted default {B}ayes factors
  for testing (in)equality constrained hypotheses}.{\BBCQ}
\newblock
\APACjournalVolNumPages{Computational Statistics and Data
  Analysis}{71}{}{448--463}.
\PrintBackRefs{\CurrentBib}

\bibitem [\protect \citeauthoryear {%
Mulder%
\ \BBA {} Fox%
}{%
Mulder%
\ \BBA {} Fox%
}{%
{\protect \APACyear {2018}}%
}]{%
Mulder:2018}
\APACinsertmetastar {%
Mulder:2018}%
\begin{APACrefauthors}%
Mulder, J.%
\BCBT {}\ \BBA {} Fox, J\BHBI P.%
\end{APACrefauthors}%
\unskip\
\newblock
\APACrefYearMonthDay{2018}{}{}.
\newblock
{\BBOQ}\APACrefatitle {Bayes factor testing of multiple intraclass
  correlations} {Bayes factor testing of multiple intraclass
  correlations}.{\BBCQ}
\newblock
\APACjournalVolNumPages{Bayesian Analysis}{14}{}{521--552}.
\PrintBackRefs{\CurrentBib}

\bibitem [\protect \citeauthoryear {%
Mulder%
, Hoijtink%
\BCBL {}\ \BBA {} de Leeuw%
}{%
Mulder%
\ \protect \BOthers {.}}{%
{\protect \APACyear {2012}}%
}]{%
Mulder:2012}
\APACinsertmetastar {%
Mulder:2012}%
\begin{APACrefauthors}%
Mulder, J.%
, Hoijtink, H.%
\BCBL {}\ \BBA {} de Leeuw, C.%
\end{APACrefauthors}%
\unskip\
\newblock
\APACrefYearMonthDay{2012}{}{}.
\newblock
{\BBOQ}\APACrefatitle {BIEMS: A {F}ortran 90 program for calculating {B}ayes
  factors for inequality and equality constrained model} {Biems: A {F}ortran 90
  program for calculating {B}ayes factors for inequality and equality
  constrained model}.{\BBCQ}
\newblock
\APACjournalVolNumPages{Journal of Statistical Software}{46}{}{}.
\PrintBackRefs{\CurrentBib}

\bibitem [\protect \citeauthoryear {%
Mulder%
, Hoijtink%
\BCBL {}\ \BBA {} Klugkist%
}{%
Mulder%
\ \protect \BOthers {.}}{%
{\protect \APACyear {2010}}%
}]{%
Mulder:2010}
\APACinsertmetastar {%
Mulder:2010}%
\begin{APACrefauthors}%
Mulder, J.%
, Hoijtink, H.%
\BCBL {}\ \BBA {} Klugkist, I.%
\end{APACrefauthors}%
\unskip\
\newblock
\APACrefYearMonthDay{2010}{}{}.
\newblock
{\BBOQ}\APACrefatitle {Equality and Inequality Constrained Multivariate Linear
  Models: Objective Model Selection Using Constrained Posterior Priors}
  {Equality and inequality constrained multivariate linear models: Objective
  model selection using constrained posterior priors}.{\BBCQ}
\newblock
\APACjournalVolNumPages{Journal of Statistical Planning and
  Inference}{140}{}{887--906}.
\PrintBackRefs{\CurrentBib}

\bibitem [\protect \citeauthoryear {%
Mulder%
\ \protect \BOthers {.}}{%
Mulder%
\ \protect \BOthers {.}}{%
{\protect \APACyear {2009}}%
}]{%
Mulder:2009}
\APACinsertmetastar {%
Mulder:2009}%
\begin{APACrefauthors}%
Mulder, J.%
, Klugkist, I.%
, van~de Schoot, A.%
, Meeus, W.%
, Selfhout, M.%
\BCBL {}\ \BBA {} Hoijtink, H.%
\end{APACrefauthors}%
\unskip\
\newblock
\APACrefYearMonthDay{2009}{}{}.
\newblock
{\BBOQ}\APACrefatitle {Bayesian Model Selection of Informative Hypotheses for
  Repeated Measurements} {Bayesian model selection of informative hypotheses
  for repeated measurements}.{\BBCQ}
\newblock
\APACjournalVolNumPages{Journal of Mathematical Psychology}{53}{}{530--546}.
\PrintBackRefs{\CurrentBib}

\bibitem [\protect \citeauthoryear {%
Mulder%
\ \BBA {} Olsson-Collentine%
}{%
Mulder%
\ \BBA {} Olsson-Collentine%
}{%
{\protect \APACyear {2019}}%
}]{%
MulderOlsson:2019}
\APACinsertmetastar {%
MulderOlsson:2019}%
\begin{APACrefauthors}%
Mulder, J.%
\BCBT {}\ \BBA {} Olsson-Collentine, A.%
\end{APACrefauthors}%
\unskip\
\newblock
\APACrefYearMonthDay{2019}{}{}.
\newblock
{\BBOQ}\APACrefatitle {Simple {B}ayesian testing of scientific expectations in
  linear regression models} {Simple {B}ayesian testing of scientific
  expectations in linear regression models}.{\BBCQ}
\newblock
\APACjournalVolNumPages{Behavioral Research Methods}{51}{}{1117--1130}.
\PrintBackRefs{\CurrentBib}

\bibitem [\protect \citeauthoryear {%
O'Hagan%
}{%
O'Hagan%
}{%
{\protect \APACyear {1995}}%
}]{%
OHagan:1995}
\APACinsertmetastar {%
OHagan:1995}%
\begin{APACrefauthors}%
O'Hagan, A.%
\end{APACrefauthors}%
\unskip\
\newblock
\APACrefYearMonthDay{1995}{}{}.
\newblock
{\BBOQ}\APACrefatitle {Fractional {B}ayes Factors for Model Comparison (with
  discussion)} {Fractional {B}ayes factors for model comparison (with
  discussion)}.{\BBCQ}
\newblock
\APACjournalVolNumPages{Journal of the Royal Statistical Society Series
  B}{57}{}{99--138}.
\PrintBackRefs{\CurrentBib}

\bibitem [\protect \citeauthoryear {%
O'Hagan%
}{%
O'Hagan%
}{%
{\protect \APACyear {1997}}%
}]{%
OHagan:1997}
\APACinsertmetastar {%
OHagan:1997}%
\begin{APACrefauthors}%
O'Hagan, A.%
\end{APACrefauthors}%
\unskip\
\newblock
\APACrefYearMonthDay{1997}{}{}.
\newblock
{\BBOQ}\APACrefatitle {Properties of intrinsic and fractional {B}ayes factors}
  {Properties of intrinsic and fractional {B}ayes factors}.{\BBCQ}
\newblock
\APACjournalVolNumPages{Test}{6}{}{101--118}.
\PrintBackRefs{\CurrentBib}

\bibitem [\protect \citeauthoryear {%
{Open Science Collaboration}%
}{%
{Open Science Collaboration}%
}{%
{\protect \APACyear {2015}}%
}]{%
OSF:2015}
\APACinsertmetastar {%
OSF:2015}%
\begin{APACrefauthors}%
{Open Science Collaboration}.%
\end{APACrefauthors}%
\unskip\
\newblock
\APACrefYearMonthDay{2015}{}{}.
\newblock
{\BBOQ}\APACrefatitle {Estimating the reproducibility of psychological science}
  {Estimating the reproducibility of psychological science}.{\BBCQ}
\newblock
\APACjournalVolNumPages{Science}{349}{}{6251}.
\PrintBackRefs{\CurrentBib}

\bibitem [\protect \citeauthoryear {%
Rouder%
, Speckman%
, D.~Sun%
\BCBL {}\ \BBA {} Iverson%
}{%
Rouder%
\ \protect \BOthers {.}}{%
{\protect \APACyear {2009}}%
}]{%
Rouder:2009}
\APACinsertmetastar {%
Rouder:2009}%
\begin{APACrefauthors}%
Rouder, J\BPBI N.%
, Speckman, P\BPBI L.%
, D.~Sun, R\BPBI D\BPBI M.%
\BCBL {}\ \BBA {} Iverson, G.%
\end{APACrefauthors}%
\unskip\
\newblock
\APACrefYearMonthDay{2009}{}{}.
\newblock
{\BBOQ}\APACrefatitle {Bayesian t tests for accepting and rejecting the null
  hypothesis} {Bayesian t tests for accepting and rejecting the null
  hypothesis}.{\BBCQ}
\newblock
\APACjournalVolNumPages{Psychonomic Bulletin \& Review}{16}{}{225--237}.
\PrintBackRefs{\CurrentBib}

\bibitem [\protect \citeauthoryear {%
Rubin%
}{%
Rubin%
}{%
{\protect \APACyear {1987}}%
}]{%
Rubin:1987}
\APACinsertmetastar {%
Rubin:1987}%
\begin{APACrefauthors}%
Rubin, D\BPBI B.%
\end{APACrefauthors}%
\unskip\
\newblock
\APACrefYear{1987}.
\newblock
\APACrefbtitle {Multiple imputation for nonresponse in surveys} {Multiple
  imputation for nonresponse in surveys}.
\newblock
\APACaddressPublisher{}{New York: John Wiley}.
\PrintBackRefs{\CurrentBib}

\bibitem [\protect \citeauthoryear {%
Rubin%
}{%
Rubin%
}{%
{\protect \APACyear {1996}}%
}]{%
Rubin:1996}
\APACinsertmetastar {%
Rubin:1996}%
\begin{APACrefauthors}%
Rubin, D\BPBI B.%
\end{APACrefauthors}%
\unskip\
\newblock
\APACrefYearMonthDay{1996}{}{}.
\newblock
{\BBOQ}\APACrefatitle {Multiple imputation after 18+ years} {Multiple
  imputation after 18+ years}.{\BBCQ}
\newblock
\APACjournalVolNumPages{Journal of the American statistical
  Association}{91}{}{473--489}.
\PrintBackRefs{\CurrentBib}

\bibitem [\protect \citeauthoryear {%
Scott%
\ \BBA {} Berger%
}{%
Scott%
\ \BBA {} Berger%
}{%
{\protect \APACyear {2006}}%
}]{%
Scott:2006}
\APACinsertmetastar {%
Scott:2006}%
\begin{APACrefauthors}%
Scott, J\BPBI G.%
\BCBT {}\ \BBA {} Berger, J\BPBI O.%
\end{APACrefauthors}%
\unskip\
\newblock
\APACrefYearMonthDay{2006}{}{}.
\newblock
{\BBOQ}\APACrefatitle {An exploration of aspects of Bayesian multiple testing}
  {An exploration of aspects of bayesian multiple testing}.{\BBCQ}
\newblock
\APACjournalVolNumPages{Journal of Statistical Planning and
  Inference}{136}{}{2144--2162}.
\PrintBackRefs{\CurrentBib}

\bibitem [\protect \citeauthoryear {%
Spiegelhalter%
\ \BBA {} Smith%
}{%
Spiegelhalter%
\ \BBA {} Smith%
}{%
{\protect \APACyear {1982}}%
}]{%
Spiegelhalter:1982}
\APACinsertmetastar {%
Spiegelhalter:1982}%
\begin{APACrefauthors}%
Spiegelhalter, D\BPBI J.%
\BCBT {}\ \BBA {} Smith, A\BPBI F\BPBI M.%
\end{APACrefauthors}%
\unskip\
\newblock
\APACrefYearMonthDay{1982}{}{}.
\newblock
{\BBOQ}\APACrefatitle {Bayes factors for linear and log-linear models with
  vague prior information} {Bayes factors for linear and log-linear models with
  vague prior information}.{\BBCQ}
\newblock
\APACjournalVolNumPages{Journal of the Royal Statistical Society Series
  B}{44}{}{377-387}.
\PrintBackRefs{\CurrentBib}

\bibitem [\protect \citeauthoryear {%
Stevens%
}{%
Stevens%
}{%
{\protect \APACyear {1996}}%
}]{%
Stevens:1996}
\APACinsertmetastar {%
Stevens:1996}%
\begin{APACrefauthors}%
Stevens, J.%
\end{APACrefauthors}%
\unskip\
\newblock
\APACrefYear{1996}.
\newblock
\APACrefbtitle {Applied Multivariate Statistics for the Social Sciences}
  {Applied multivariate statistics for the social sciences}.
\newblock
\APACaddressPublisher{}{Mahwah NJ: Lawrence Erlbaum}.
\PrintBackRefs{\CurrentBib}

\bibitem [\protect \citeauthoryear {%
{Van de Schoot}%
\ \protect \BOthers {.}}{%
{Van de Schoot}%
\ \protect \BOthers {.}}{%
{\protect \APACyear {2006}}%
}]{%
Schoot:2011c}
\APACinsertmetastar {%
Schoot:2011c}%
\begin{APACrefauthors}%
{Van de Schoot}, R.%
, Hoijtink, H.%
, Mulder, J.%
, Aken, M.%
, {de Castro}, B.%
, Meeus, W.%
\BCBL {}\ \BBA {} Romeijn, J\BHBI W.%
\end{APACrefauthors}%
\unskip\
\newblock
\APACrefYearMonthDay{2006}{}{}.
\newblock
{\BBOQ}\APACrefatitle {Evaluating expectations about negative emotional states
  of aggressive boys using {B}ayesian model selection} {Evaluating expectations
  about negative emotional states of aggressive boys using {B}ayesian model
  selection}.{\BBCQ}
\newblock
\APACjournalVolNumPages{Developmental Psychology}{47}{}{203--212}.
\PrintBackRefs{\CurrentBib}

\bibitem [\protect \citeauthoryear {%
{van Schie}%
, {Van Veen}%
, Engelhard%
, Klugkist%
\BCBL {}\ \BBA {} {Van den Hout}%
}{%
{van Schie}%
\ \protect \BOthers {.}}{%
{\protect \APACyear {2016}}%
}]{%
vanSchie:2016}
\APACinsertmetastar {%
vanSchie:2016}%
\begin{APACrefauthors}%
{van Schie}, K.%
, {Van Veen}, S.%
, Engelhard, I.%
, Klugkist, I.%
\BCBL {}\ \BBA {} {Van den Hout}, M.%
\end{APACrefauthors}%
\unskip\
\newblock
\APACrefYearMonthDay{2016}{}{}.
\newblock
{\BBOQ}\APACrefatitle {Blurring emotional memories using eye movements:
  Individual differences and speed of eye movements} {Blurring emotional
  memories using eye movements: Individual differences and speed of eye
  movements}.{\BBCQ}
\newblock
\APACjournalVolNumPages{European Journal of Psychotraumatology}{7}{}{}.
\PrintBackRefs{\CurrentBib}

\bibitem [\protect \citeauthoryear {%
Vrinten%
\ \protect \BOthers {.}}{%
Vrinten%
\ \protect \BOthers {.}}{%
{\protect \APACyear {2016}}%
}]{%
Vrinten:2016}
\APACinsertmetastar {%
Vrinten:2016}%
\begin{APACrefauthors}%
Vrinten, C.%
, Gu, X.%
, %
, Weinreich, S.%
, Schipper, M.%
, Wessels, J.%
\BDBL {}Verschuuren, J.%
\end{APACrefauthors}%
\unskip\
\newblock
\APACrefYearMonthDay{2016}{}{}.
\newblock
{\BBOQ}\APACrefatitle {An n-of-one RCT for intravenous immunoglobulin G for
  inflammation in hereditary neuropathy with liability to pressure palsy
  (HNPP)} {An n-of-one rct for intravenous immunoglobulin g for inflammation in
  hereditary neuropathy with liability to pressure palsy (hnpp)}.{\BBCQ}
\newblock
\APACjournalVolNumPages{Journal of Neurology, Neurosurgery and
  Psychiatry}{87}{}{790--791}.
\PrintBackRefs{\CurrentBib}

\bibitem [\protect \citeauthoryear {%
Well%
, Kolk%
\BCBL {}\ \BBA {} Klugkist%
}{%
Well%
\ \protect \BOthers {.}}{%
{\protect \APACyear {2008}}%
}]{%
VanWell:2008}
\APACinsertmetastar {%
VanWell:2008}%
\begin{APACrefauthors}%
Well, S\BPBI V.%
, Kolk, A.%
\BCBL {}\ \BBA {} Klugkist, I.%
\end{APACrefauthors}%
\unskip\
\newblock
\APACrefYearMonthDay{2008}{}{}.
\newblock
{\BBOQ}\APACrefatitle {Effects of Sex, Gender Role Identification, and Gender
  relevance of Two Types of Stressors on Cardiovascular and Subjective
  Responses: Sex and Gender Match/Mismatch Effects} {Effects of sex, gender
  role identification, and gender relevance of two types of stressors on
  cardiovascular and subjective responses: Sex and gender match/mismatch
  effects}.{\BBCQ}
\newblock
\APACjournalVolNumPages{Behavior Modification}{32}{}{427--449}.
\PrintBackRefs{\CurrentBib}

\bibitem [\protect \citeauthoryear {%
Zondervan-Zwijnenburg%
\ \protect \BOthers {.}}{%
Zondervan-Zwijnenburg%
\ \protect \BOthers {.}}{%
{\protect \APACyear {2019}}%
}]{%
Zondervan:2019}
\APACinsertmetastar {%
Zondervan:2019}%
\begin{APACrefauthors}%
Zondervan-Zwijnenburg, M.%
, Veldkamp, S.%
, Neumann, A.%
, Barzeva, S.%
, Nelemans, S.%
, {van Beijsterveldt}, C.%
\BDBL {}Boomsma, A\BPBI O\BPBI D.%
\end{APACrefauthors}%
\unskip\
\newblock
\APACrefYearMonthDay{2019}{}{}.
\newblock
{\BBOQ}\APACrefatitle {Parental Age and Offspring Childhood Mental Health: A
  Multi-Cohort, Population-Based Investigation} {Parental age and offspring
  childhood mental health: A multi-cohort, population-based
  investigation}.{\BBCQ}
\newblock
\APACjournalVolNumPages{Child Development}{}{}{}.
\PrintBackRefs{\CurrentBib}

\end{thebibliography}

\appendix

\section{Proof of Lemma 1}
The constrained model in \eqref{Mt} can equivalently be written in the parameterization $\bm\zeta_t=(\bm\zeta_{t,E},\bm\zeta_{t,O})'=\textbf{H}_t\bm\theta$, where $\textbf{H}_t=[\textbf{R}_{t,E}' ~\textbf{D}_t' ]'$ and $\textbf{D}_t$ is a $(PK-r_{r,t})\times PK$ matrix with independent rows of the form $(1,0,\ldots,0)$ with a 1 in the $PK-r_{r,t}$ columns that correspond to parameters that are not constrained with an equality constraint, such the transformation is one-to-one, as $M_t:\bm\zeta_{t,E}=\textbf{r}_{t,E}~\&~\tilde{\textbf{R}}_{t,O}\bm\zeta_{t,O}>\textbf{r}_{t,O}$, with $\tilde{\textbf{R}}_{t,O}$ are the columns of $\textbf{R}_{t,O}$ of the parameters that are not equality constrained. Furthermore, the adjusted parameter space in the denominator in \eqref{marglikeMt3} becomes
\begin{align*}
&\mathcal{M}_t^*=\{\bm\zeta_t| \bm\zeta_{t,E}-\hat{\bm\zeta}_{t,E}=\textbf{0},~\tilde{\textbf{R}}_{t,O}(\bm\zeta_{t,O}-\hat{\bm\zeta}_{t,O})>\textbf{0}\},
\end{align*}\\
%
where $\hat{\bm\zeta}_{t,E}=\textbf{R}_{t,E}\hat{\bm\theta}$, and $\hat{\bm\zeta}_{t,O}=\textbf{D}_{t}\hat{\bm\theta}$.
The marginal likelihood under $M_t$ in \eqref{marglikeMt3} can then equivalently be written as
\begin{eqnarray*}
p^*_t(\textbf{Y},\textbf{b}) &=& \frac{\int_{\bm\Sigma}\int_{\bm{\zeta}_{t,O}\in\mathcal{M}_t} p(\textbf{Y}|\textbf{X},\bm{\zeta}_{t,E}=\textbf{r}_{t,E},\bm\zeta_{t,O},\bm\Sigma)|\bm\Sigma|^{-\frac{P+1}{2}}d\bm\zeta_{t,O}d\bm\Sigma}
{\int_{\bm\Sigma}\int_{\bm\zeta_{t,O}\in\mathcal{M}^*_t} p(\textbf{Y}|\textbf{X},\bm{\zeta}_{t,E}=\hat{\bm{\zeta}}_{t,E},\bm\zeta_{t,O},\bm\Sigma)^{\textbf{b}}|\bm\Sigma|^{-\frac{P+1}{2}}d\bm\zeta_{t,O}d\bm\Sigma},
\end{eqnarray*}
and the marginal likelihood under an unconstrained alternative model, $M_u$, equals
\begin{eqnarray*}
p_u(\textbf{Y},\textbf{b}) &=& \frac{\int_{\bm\Sigma}\int_{\bm{\zeta}_{t}} p(\textbf{Y}|\textbf{X},\bm{\zeta}_{t},\bm\Sigma)|\bm\Sigma|^{-\frac{P+1}{2}}d\bm{\zeta}_{t}d\bm\Sigma}
{\int_{\bm\Sigma}\int_{\bm{\zeta}_{t}} p(\textbf{Y}|\textbf{X},\bm{\zeta}_{t},\bm\Sigma)^{\textbf{b}}|\bm\Sigma|^{-\frac{P+1}{2}}d\bm{\zeta}_{t}d\bm\Sigma}.
\end{eqnarray*} 
Thus, the Bayes factor can be written as
\begin{eqnarray}
\nonumber B_{1u,b} &=& \frac{p_1^*(\textbf{Y},\textbf{b})}{p_u(\textbf{Y},\textbf{b})}
= \frac{\int_{\bm\Sigma}\int_{\bm{\zeta}_{t,O}\in\mathcal{M}_t} p(\textbf{Y}|\textbf{X},\bm{\zeta}_{t,E}=\textbf{r}_{t,E},\bm\zeta_{t,O},\bm\Sigma)|\bm\Sigma|^{-\frac{P+1}{2}}d\bm\zeta_{t,O}d\bm\Sigma}
{\int_{\bm\Sigma}\int_{\bm\zeta_{t,O}\in\mathcal{M}_t^*} p(\textbf{Y}|\textbf{X},\bm{\zeta}_{t,E}=\hat{\bm\zeta}_{t,E},\bm\zeta_{t,O},\bm\Sigma)^{\textbf{b}}|\bm\Sigma|^{-\frac{P+1}{2}}d\bm\zeta_{t,O}d\bm\Sigma}\text{\Huge{/}}\\
\nonumber && \frac{\int_{\bm\Sigma}\int_{\bm{\zeta}_{t}} p(\textbf{Y}|\textbf{X},\bm{\zeta}_{t},\bm\Sigma)|\bm\Sigma|^{-\frac{P+1}{2}}d\bm{\zeta}_{t}d\bm\Sigma}
{\int_{\bm\Sigma}\int_{\bm{\zeta}_{t}} p(\textbf{Y}|\textbf{X},\bm{\zeta}_{t},\bm\Sigma)^{\textbf{b}}|\bm\Sigma|^{-\frac{P+1}{2}}d\bm{\zeta}_{t}d\bm\Sigma}\\
\nonumber &=&
\int_{\bm\Sigma}\int_{\bm{\zeta}_{t,O}\in\mathcal{M}_t}
\frac{p(\textbf{Y}|\textbf{X},\bm{\zeta}_{t,E}=\textbf{r}_{t,E},\bm\zeta_{t,O},\bm\Sigma)|\bm\Sigma|^{-\frac{P+1}{2}}}
{\int_{\bm\Sigma}\int_{\bm{\zeta}_{t}} p(\textbf{Y}|\textbf{X},\bm{\zeta}_{t},\bm\Sigma)|\bm\Sigma|^{-\frac{P+1}{2}}d\bm{\zeta}_{t}d\bm\Sigma}d\bm\zeta_{t,O}d\bm\Sigma\text{\Huge{/}}\\
\nonumber&&\int_{\bm\Sigma}\int_{\bm\zeta_{t,O}\in\mathcal{M}_t^*}
\frac{p(\textbf{Y}|\textbf{X},\bm{\zeta}_{t,E}=\hat{\bm\zeta}_{t,E},\bm\zeta_{t,O},\bm\Sigma)^{\textbf{b}}|\bm\Sigma|^{-\frac{P+1}{2}}}
{\int_{\bm\Sigma}\int_{\bm{\zeta}_{t}}p(\textbf{Y}|\textbf{X},\bm{\zeta}_{t},\bm\Sigma)^{\textbf{b}}|\bm\Sigma|^{-\frac{P+1}{2}}d\bm\zeta_{t,O}d\bm\Sigma}d\bm{\zeta}_{t}d\bm\Sigma\\
\nonumber&=& \int_{\bm\Sigma}\int_{\bm{\zeta}_{t,O}\in\mathcal{M}_t}
\pi_u(\bm{\zeta}_{t,E}=\textbf{r}_{t,E},\bm\zeta_{t,O},\bm\Sigma|\textbf{Y},\textbf{X})
d\bm{\zeta}_{t}d\bm\Sigma\text{\Huge{/}}\\
\nonumber&&
\int_{\bm\Sigma}\int_{\bm{\zeta}_{t,O}\in\mathcal{M}_t^*}
\pi_u(\bm{\zeta}_{t,E}=\hat{\bm\zeta}_{t,E},\bm\zeta_{t,O},\bm\Sigma|\textbf{Y},\textbf{X},\textbf{b})
d\bm{\zeta}_{t}d\bm\Sigma\\
\nonumber&=& \int_{\bm{\zeta}_{t,O}\in\mathcal{M}_t}
\pi_u(\bm{\zeta}_{t,E}=\textbf{r}_{t,E},\bm\zeta_{t,O}|\textbf{Y},\textbf{X})
d\bm{\zeta}_{t}\text{\Huge{/}}
\int_{\bm{\zeta}_{t,O}\in\mathcal{M}_t}
\pi^*_u(\bm{\zeta}_{t,E}=\textbf{r}_{t,E},\bm\zeta_{t,O}|\textbf{Y},\textbf{X},\textbf{b})
d\bm{\zeta}_{t}\\
\label{BFsteps}&=& \frac{\pi_u(\bm{\zeta}_{t,E}=\textbf{r}_{t,E}|\textbf{Y},\textbf{X})}
{\pi_u^*(\bm{\zeta}_{t,E}=\textbf{r}_{t,E}|\textbf{Y},\textbf{X},\textbf{b})}
\times
\frac{\text{Pr}_u(\tilde{\textbf{R}}_{t,O}\bm\zeta_{t,O}>\tilde{\textbf{r}}_{t,O}|\bm{\zeta}_{t,E}=\textbf{r}_{t,E},\textbf{Y},\textbf{X})}
{\text{Pr}^*_u(\tilde{\textbf{R}}_{t,O}\bm\zeta_{t,O}>\tilde{\textbf{r}}_{t,O}|\bm{\zeta}_{t,E}=\textbf{r}_{t,E},\textbf{Y},\textbf{X},\textbf{b})},
\end{eqnarray}
where
\[
\pi_u^*(\bm{\zeta}_{t,E},\bm{\zeta}_{t,O}|\textbf{Y},\textbf{X},\textbf{b})=\pi_u(\bm{\zeta}_{t,E}+\hat{\bm{\zeta}}_{t,E}-\textbf{r}_{t,E},\bm{\zeta}_{t,O}+\hat{\bm{\zeta}}_{t,O}-\bm{\zeta}_{t,O,0}|\textbf{Y},\textbf{X},\textbf{b}).
\]

The unconstrained marginal and conditional posteriors follow naturally from Bayes' theorem,
\begin{eqnarray}
\nonumber \pi_u(\bm\theta,\bm\Sigma|\textbf{Y},\textbf{X}) &\propto & |\bm\Sigma|^{-\frac{P+1}{2}}p(\textbf{Y}|\textbf{X},\bm{\Theta},\bm\Sigma)\\
\nonumber &\propto & |\bm\Sigma|^{-\frac{N+P+1}{2}}\exp\{
-\tfrac{1}{2}\text{tr}~\bm\Sigma^{-1}({\textbf{Y}}-{\textbf{X}}\bm{\Theta})'({\textbf{Y}}-{\textbf{X}}\bm{\Theta})
\}\\
\nonumber &\propto& \pi_u(\bm{\Theta}|\textbf{Y},\textbf{X},\bm\Sigma)\pi_u(\bm\Sigma|\textbf{Y},\textbf{X}),\\
\text{with }\pi(\bm{\Theta}|\textbf{Y},\textbf{X},\bm\Sigma)&= & \mathcal{N}_{K,P}(\hat{\bm{\Theta}},({\textbf{X}}'{\textbf{X}})^{-1},\bm\Sigma)\\
\pi(\bm\Sigma|\textbf{Y},\textbf{X})&=&
\mathcal{IW}(N-K,{\textbf{S}})
\end{eqnarray}
where the least squares estimate is given by $\hat{\bm{\Theta}}=(\textbf{X}'{\textbf{X}})^{-1}{\textbf{X}}'{\textbf{Y}}$ and the sums of square matrix equals ${\textbf{S}}=({\textbf{Y}} - {\textbf{X}}\hat{{\bm{\Theta}}})'({\textbf{Y}} - {\textbf{X}}\hat{{\bm{\Theta}}})$. Furthermore, $\mathcal{N}_{K,P}$ and $\mathcal{IW}$ denote a matrix normal distribution for a $K\times P$ matrix and an inverse Wishart distribution, respectively. Note that the conditional posterior distribution for $\bm{\Theta}$ is equivalent to a multivariate normal on the vectorization, $\pi(\bm\theta|\textbf{Y},\textbf{X},\bm\Sigma) = \mathcal{N}(\hat{\bm\theta},\bm\Sigma\otimes({\textbf{X}}'{\textbf{X}})^{-1})$. Integrating the covariance matrix out results in a marginal posterior for $\bm{\Theta}$ having a $K\times P$ matrix Student $t$ distribution, 
\[
\pi(\bm{\Theta}|\textbf{Y},\textbf{X}) = \mathcal{T}_{K,P}(
\hat{\bm{\Theta}}, ({\textbf{X}}'{\textbf{X}})^{-1}, {\textbf{S}},N-K-P+1).
\]

The unconstrained default prior is obtained by first raising the likelihood of the $i$-th observation to a fraction $b_i$, i.e.,
\begin{eqnarray*}
p(\textbf{y}_i|\textbf{x}_i,\bm{\Theta},\bm\Sigma)^{b_i} &\propto &
|\bm\Sigma|^{-\frac{b_i}{2}}\exp\{
-\tfrac{b_i}{2}(\textbf{y}_i-\bm{\Theta}'\textbf{x}_i)'\bm\Sigma^{-1}
(\textbf{y}_i-\bm{\Theta}'\textbf{x}_i)
\}\\
&= &
|\bm\Sigma|^{-\frac{b_i}{2}}\exp\{
-\tfrac{1}{2}({\textbf{y}}_{i,b_i}-\bm{\Theta}'{\textbf{x}}_{i,b_i})'\bm\Sigma^{-1}
({\textbf{y}}_{i,b_i}-\bm{\Theta}'{\textbf{x}}_{i,b_i})
\},
\end{eqnarray*}
where ${\textbf{y}}_{i,b_i}=\sqrt{b_i}\textbf{y}_i$ and ${\textbf{x}}_{i,b_i}=\sqrt{b_i}\textbf{x}_i$. The likelihood raised to observation specific fractions is then defined as
\begin{eqnarray*}
p(\textbf{Y}|\textbf{X},\bm{\Theta},\bm\Sigma)^{\textbf{b}} &\equiv&\prod_{i=1}^np(\textbf{y}_i|\textbf{x}_i,\bm{\Theta},\bm\Sigma)^{b_i} \\
&\propto&
|\bm\Sigma|^{-\frac{1}{2}\sum_{i=1}^nb_i}\exp\{
-\tfrac{1}{2}\text{tr}~\bm\Sigma^{-1}({\textbf{Y}}_{\textbf{b}}-{\textbf{X}}_{\textbf{b}}\bm{\Theta})'({\textbf{Y}}_{\textbf{b}}-{\textbf{X}}_{\textbf{b}}\bm{\Theta})
\}\\
&=& |\bm\Sigma|^{-\frac{1}{2}\sum_{i=1}^nb_i}\exp\{
-\tfrac{1}{2}\text{tr}~\bm\Sigma^{-1}{\textbf{S}}_{\textbf{b}}\}\\
&&\exp\{ -\tfrac{1}{2}\text{tr}~\bm\Sigma^{-1}(\bm{\Theta}-\hat{{\bm{\Theta}}}_{\textbf{b}})'{\textbf{X}}'_{\textbf{b}}{\textbf{X}}_{\textbf{b}}(\bm{\Theta}-\hat{{\bm{\Theta}}}_{\textbf{b}})
\},
\end{eqnarray*}
where the least squares estimate is given by $\hat{{\bm{\Theta}}}_{\textbf{b}}=({\textbf{X}}'_{\textbf{b}}{\textbf{X}}_{\textbf{b}})^{-1}{\textbf{X}}_{\textbf{b}}'{\textbf{Y}}_{\textbf{b}}$ and the sums of square matrix equals ${\textbf{S}}_{\textbf{b}}=({\textbf{Y}}_{\textbf{b}} - {\textbf{X}}_{\textbf{b}}\hat{{\bm{\Theta}}}_{\textbf{b}})'({\textbf{Y}}_{\textbf{b}} - {\textbf{X}}_{\textbf{b}}\hat{{\bm{\Theta}}}_{\textbf{b}})$, and $\textbf{Y}_{\textbf{b}}$ and $\textbf{X}_{\textbf{b}}$ are the stacked matrices of $\textbf{y}_{i,b_i}'$ and $\textbf{x}_{i,b_i}'$, respectively. In combination with the improper noninformative independence Jeffreys' prior, the fractional default prior based on generalized fractional Bayes methodology can then be written as
\begin{eqnarray*}
\nonumber \pi_u(\bm{\Theta},\bm\Sigma|\textbf{Y},\textbf{X},{\textbf{b}}) \nonumber &\propto& |\bm\Sigma|^{-\frac{P+1}{2}}p(\textbf{Y}|\textbf{X},\bm{\Theta},\bm\Sigma)^{\textbf{b}}\\
\nonumber &\propto & \pi(\bm{\Theta}|\bm\Sigma,\textbf{Y},\textbf{X},{\textbf{b}})\pi_u(\bm\Sigma|\textbf{Y},\textbf{X},{\textbf{b}}),\\
\text{with }\pi_u(\bm{\Theta}|\bm\Sigma,\textbf{Y},\textbf{X},{\textbf{b}})&= & \mathcal{N}_{K,P}(\hat{{\bm{\Theta}}}_{\textbf{b}},({\textbf{X}}'_{\textbf{b}}{\textbf{X}}_{\textbf{b}})^{-1},\bm\Sigma),\\
\nonumber \pi_u(\bm\Sigma|\textbf{Y},\textbf{X},{\textbf{b}})&=&
\mathcal{IW}(\sum_{i=1}^n b_i-K,{\textbf{S}}_{\textbf{b}}),
\end{eqnarray*}
so that
\begin{eqnarray*}
\pi_u(\bm{\Theta}|\textbf{Y},\textbf{X},{\textbf{b}}) &=& \mathcal{T}_{K,P}(
\hat{{\bm{\Theta}}}_{\textbf{b}}, {\textbf{S}}_{\textbf{b}}, ({\textbf{X}}'_{\textbf{b}}{\textbf{X}}_{\textbf{b}})^{-1},\sum_{i=1}^N b_i-K-P+1).
\end{eqnarray*}

Finally, integrating the unconstrained prior $\pi_u$ over the adjusted subspace $\mathcal{M}_t^*$ in step 4 of \eqref{BFsteps} is equivalent to integrating adjusted unconstrained priors $\pi^*$ over $\mathcal{M}_t$,
\begin{eqnarray*}
\pi^*_u(\bm{\Theta}|\bm\Sigma,\textbf{Y},\textbf{X},{\textbf{b}})&= & \mathcal{N}_{K,P}(\bm{\Theta}_0,({\textbf{X}}'_{\textbf{b}}{\textbf{X}}_{\textbf{b}})^{-1},\bm\Sigma)\\
\nonumber \pi^*_u(\bm\Sigma|\textbf{Y},\textbf{X},{\textbf{b}})&=&
\mathcal{IW}(\sum_{i=1}^N b_i-K,{\textbf{S}}_{\textbf{b}}),\\
\pi^*_u(\bm{\Theta}|\textbf{Y},\textbf{X},{\textbf{b}}) &=& \mathcal{T}_{K,P}(\sum_{i=1}^N b_i-K-P+1,
\bm{\Theta}_0, {\textbf{S}}_{\textbf{b}}, ({\textbf{X}}'_{\textbf{b}}{\textbf{X}}_{\textbf{b}})^{-1}).
\end{eqnarray*}

\end{document}